\documentclass[aps,pre,twocolumn,groupedaddress,showpacs,superscriptaddress,amssymb,amsmath]{revtex4-2}
\usepackage{graphicx}
\usepackage{dcolumn}
\usepackage{bm}
\usepackage{hyperref}
\usepackage{cleveref}
\hypersetup{
    colorlinks=true,
    linkcolor=blue,
	citecolor=cyan
}
\usepackage{comment}
\usepackage{color}

\usepackage[utf8]{inputenc}
\usepackage{graphicx}
\usepackage{tabularx}
\usepackage{xcolor}
\usepackage{amsmath}
\usepackage{dcolumn}
\usepackage{hyperref}
\usepackage{bm}
\usepackage{epsf}
\usepackage{braket}
\usepackage{tensor}
\usepackage{soul}

\begin{document}
\newcolumntype{M}[1]{>{\centering\arraybackslash}m{#1}}

\title{Filling an empty lattice by local injection of quantum particles}

\author{Akash Trivedi}
\email{akash.trivedi@students.iiserpune.ac.in}
\affiliation{Department of Physics, Indian Institute of Science Education and Research Pune, Dr. Homi Bhabha Road, Ward No. 8, NCL Colony, Pashan, Pune, Maharashtra 411008, India}
\affiliation{International Centre for Theoretical Sciences, Tata Institute of Fundamental Research,
Bangalore 560089, India}

\author{Bijay Kumar Agarwalla}
\email{bijay@iiserpune.ac.in}
\affiliation{Department of Physics, Indian Institute of Science Education and Research Pune, Dr. Homi Bhabha Road, Ward No. 8, NCL Colony, Pashan, Pune, Maharashtra 411008, India}

\author{Abhishek Dhar}
\email{abhishek.dhar@icts.res.in}
\affiliation{International Centre for Theoretical Sciences, Tata Institute of Fundamental Research,
Bangalore 560089, India}

\author{Manas Kulkarni}
\email{manas.kulkarni@icts.res.in} 
\affiliation{International Centre for Theoretical Sciences, Tata Institute of Fundamental Research,
Bangalore 560089, India}

\author{Anupam Kundu}
\email{anupam.kundu@icts.res.in} 
\affiliation{International Centre for Theoretical Sciences, Tata Institute of Fundamental Research,
Bangalore 560089, India}

\author{Sanjib Sabhapandit}
\email{sanjib@rri.res.in} 
\affiliation{Raman Research Institute, Bangalore 560080, India}

\date{\today}

\begin{abstract}
We study the  quantum dynamics of filling an empty lattice of size $L$, by connecting it locally with an equilibrium thermal bath that injects non-interacting bosons or fermions. We adopt four different approaches, namely (i)~direct exact numerics, (ii)~Redfield equation, (iii)~Lindblad equation, and (iv)~quantum Langevin equation ---  which are unique in their ways for solving the time dynamics and the steady-state. Our setup offers a simplistic platform to understand fundamental aspects of dynamics and approach to thermalization. The quantities of interest that we consider are the spatial density profile and the total number of bosons/fermions in the lattice. The spatial spread is ballistic in nature and the local occupation eventually settles down owing to equilibration. 
The ballistic spread of local density admits a universal scaling form. We show that this universality is  only seen when the condition of detailed balance is satisfied by the baths.  The difference between bosons and fermions shows up in the early time growth rate and the saturation values of the profile. The techniques developed here are applicable to systems in arbitrary dimensions and for arbitrary geometries. 
\end{abstract}

\maketitle


\section{Introduction\label{sec:Introduction}}
Understanding quantum dynamics and subsequent thermalization of a system in presence of a bath is an interesting question in open quantum systems both from a fundamental and an applied perspective \cite{breuer2002theory, carmichael2009statistical,rotter2015review,rivas2012open,RevModPhys.88.021002,weiss2012quantum}. In this regard, a plethora of studies have emerged in successfully addressing some of the pressing issues \cite{Thermal-OQS,Thermo-Dario}. A good starting point of such an investigation is to understand the quantum dynamics and subsequent equilibration that an empty lattice would undergo when attached to a reservoir. An intricate interplay between hermitian, non-hermitian dynamics (arising due to finite system-reservoir coupling) and quantum statistics can lead to a highly non-trivial dynamics and steady-state. 

In this direction, quantum dynamics involving localized source/sink has been an active area of research \cite{froml2020ultracold, barmettler2011controllable,zezyulin2012macroscopic,
kordas2013decay,kiefer2017current,sels2020thermal,labouvie2016bistability,mullers2018coherent, krapivsky2019free,krapivsky2020free,butz2010dynamical,alba2021noninteracting,krapivsky2014survival}. Given the complexity of such setups, one is often compelled to resort to approximations such as: weak system-reservoir coupling and separation of time-scales between system and reservoir dynamics. Albeit quite successful \cite{krapivsky2019free,krapivsky2020free}, these approximations might miss certain key aspects of quantum dynamics and thermalization. For example, a non-monotonic behaviour of out-of-equilibrium transport properties when one tunes the system-reservoir coupling from weak to strong may be missed in traditional perturbative approaches. Furthermore, the assumption of well separated reservoir and system time scales can become invalid for a wide class of baths with spectral functions of non-analytical type \cite{PhysRevB.97.104306}. Therefore it is crucial to employ exact approaches to investigate such setups.

 \begin{figure}[t]
     \includegraphics[width=0.4\textwidth]{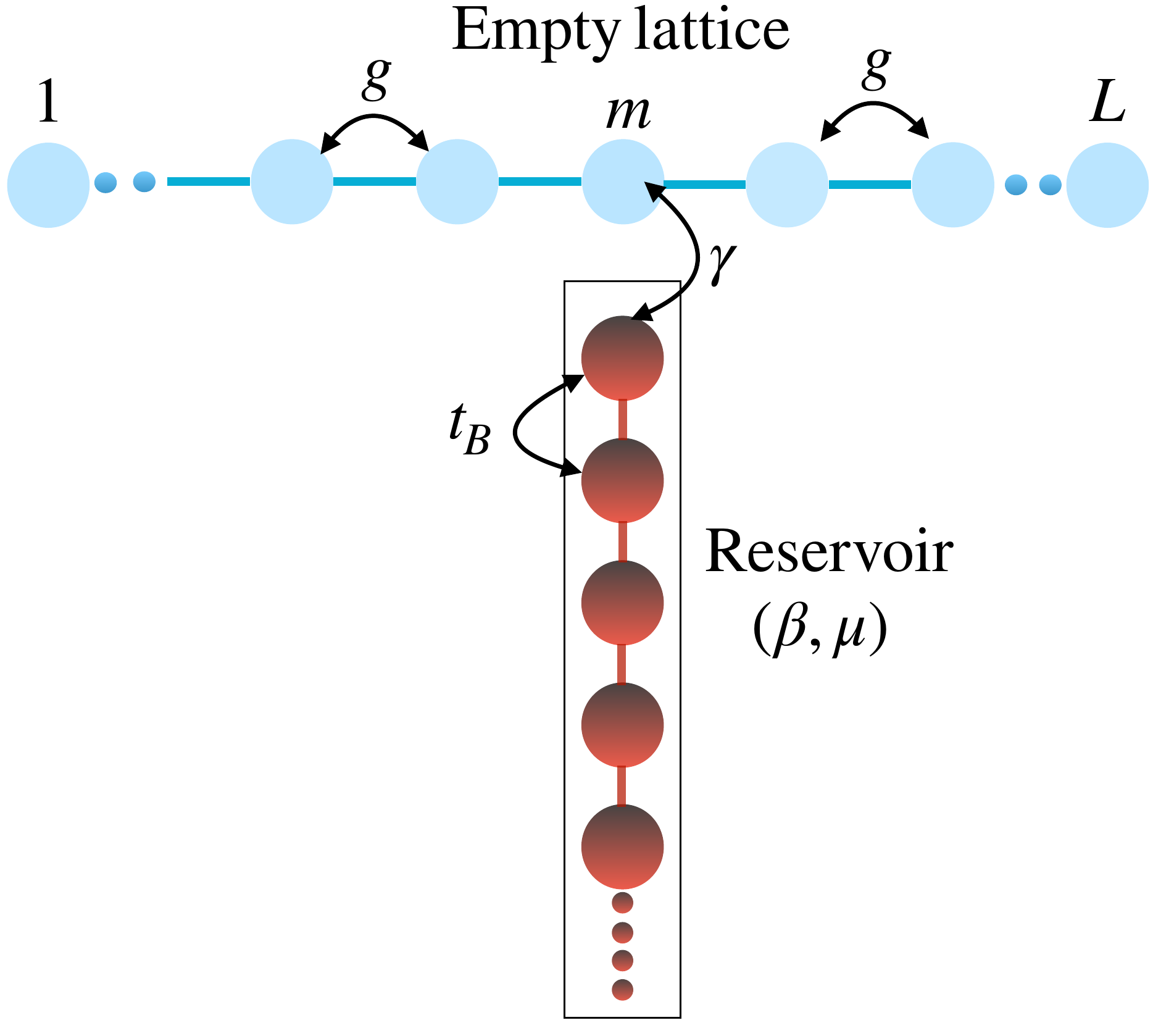}
     \caption{Schematic of our setup: an empty (blue) one-dimensional lattice of size $L$ is connected to a reservoir (red) at a particular site. The reservoir is coupled to the lattice at a particular site, represented by $m$. The inter-site hopping parameter within the lattice (within the reservoir) is $g$ ($t_B$). The coupling between the lattice and the reservoir is denoted by $\gamma$. The reservoir is maintained in equilibrium at an inverse temperature $\beta$ and chemical potential $\mu$.}
     \label{schematic}
 \end{figure}

In this work, we consider a one-dimensional empty lattice (system) of size $L$ coupled to a boson/fermion reservoir at a particular site. Some natural questions that come to mind are: how many particles are there in the lattice at a particular time ? What is the space resolved density profile at any given time snapshot ? How does the density profile spreads with time ?  In our work we study quantum dynamics of (i) local density profile on the lattice, $n_i(t)$ and (ii) total number of bosons/fermions on the lattice, $N(t) = \sum_{i=1}^{L} n_i(t)$, and their subsequent thermalization (or lack thereof) properties. A schematic of our setup is given in Fig.~(\ref{schematic}). We employ four methods which are unique in their own ways and offer complementary benefits-- (i) direct exact numerics for correlation matrix, (ii) Redfield equation (perturbative and Markovian), (iii) local Lindblad equation (perturbative,  Markovian, and weak inter-site hopping), and (iv) exact quantum Langevin equation. The summary of our work is as follows:
\begin{enumerate}
    \item The initial growth for the total occupation $N(t)$ for bosons/fermions is linear in time and it subsequently saturates (for a finite lattice size) to a constant value in an exponential fashion [Fig.~(\ref{f2})]. For infinite lattice there is no saturation and $N(t) \propto t \, \, \forall\,\, t$ [Fig.~(\ref{f3})].
    
    \item The local spatial density profile $n_i(t)$ exhibits a ballistic spread for both bosons [Fig.~(\ref{f4})] and fermions. For finite lattice, at a given site, $n_i(t)$ initially grows in time and eventually saturates owing to equilibration with the bath.
    
    \item We observe much slower accumulation of fermions in comparison to bosons which is rooted in quantum statistics (Pauli exclusion principle). Albeit there is an analogy between fermions and classical exclusion processes, there are interesting differences \cite{krapivsky2012symmetric,krapivsky2014lattice} for the case of fermions which are intrinsically quantum in nature. For example, the spread of the density profile in the classical exclusion case is diffusive in contrast to ballistic spread for the quantum case.  
    
    \item Our microscopic starting point is drastically different from the phenomenological approaches  such as unidirectional filling \cite{krapivsky2019free,krapivsky2020free} and therefore lacking the detailed balance condition. In our work, we argue that this detailed balance condition plays a paramount role in deciding the fate of the dynamics. Nonetheless, the techniques developed in Refs.~(\onlinecite{krapivsky2019free,krapivsky2020free}) can be adapted to obtain analytical forms for spatial density profile [Fig.~(\ref{f8}), and Fig.~(\ref{f10})]. 
    
    \item In a suitable parameter regime, we find that the spatial density profile in the ballistic regime possesses a universal scaling form [Eqs.~\eqref{eq:ni_scale-main-1} and \eqref{eq:ni_scale-main-2}], valid for both bosons and fermions. 
    
\end{enumerate}
 
The plan of the paper is as follows: In Section \ref{sec:setup}, we first introduce the setup and discuss the four methods. In Section \ref{sec:numerical_results}, we present our findings for bosons using all the four methods and highlight similarities and differences for fermions. In Section \ref{sec:comparison}, we place our work in the context of recent works and provide a detailed comparison. Finally, we summarize our results along with an outlook in Section \ref{sec:Conclusions}.  We delegate some details to the appendix.

\section{Setup and Methodology}
\label{sec:setup}
In this section, we discuss our microscopic model for the injection of identical bosons/fermions in a one-dimensional lattice. The lattice  initially is an empty tight-binding chain of $L$ sites. The Hamiltonian of the lattice is given by (setting $\hbar=1$ throughout the paper), 
\begin{equation}
\label{eq:hs}
H_{S} =  g \sum_{i = 1}^{L - 1} \left({a}^{\dagger}_{i} {a}_{i+1} + {a}^{\dagger}_{i+1} {a}_{i}\right)
\end{equation}
where ${a}_{i}$ (${a}_i^{\dagger})$ is the annihilation (creation) operator of the $i$-th site of the lattice. The hopping parameter is denoted by $g$. We inject the particles at a particular site (for example, near the middle) of the lattice by putting it in contact with a reservoir which is represented by a semi-infinite tight-binding chain whose Hamiltonian is given by,
\begin{equation}
\label{eq:bath}
H_{B} = t_{B} \sum_{i = 1}^{\infty}\left({b}^{\dagger}_{i}  {b}_{i+1} +  {b}^{\dagger}_{i+1}  {b}_{i}\right).
\end{equation}
Here ${b}_{i}$ (${b}_i^{\dagger})$ is the annihilation (creation) operator of the $i$-th site of the bath and $t_B$ is the nearest neighbour hopping between the bath sites. Note that $a_i$ and $b_i$ operators satisfy either commutation or anti-commutation algebra for bosons or fermions, respectively. At $t = 0$, we switch on the coupling between the lattice and the bath which can be described by the following Hamiltonian: 
\begin{equation}
\label{LIQP2}
      {H}_{SB} = \gamma \left(  {a}^{\dagger}_{m}   {b}_{1} +   {b}^{\dagger}_{1}   {a}_{m} \right)
\end{equation}
where $\gamma$ is the coupling strength and the $m$-th site of the lattice is coupled to the first site of the bath. In this work we choose $m=(N+1)/2$ for odd $N$ and $m=N/2+1$ for even $N$. The initial density operator $\rho(t=0)$ for the setup is taken as a product state
\begin{equation}
\label{eq:globalrho}
\rho(0)= \rho_S(0) \otimes \rho^{\rm th}_{B}(0)
\end{equation} 
with the lattice initially being empty i.e., $\rho_S(t\!=\!0)\!=\! |\bf{0}\rangle \langle \bf{0}|$ where $|\bf{0}\rangle$ denotes a state with all sites empty. In other words, the lattice is in vacuum. The bath density operator is in equilibrium at temperature $T = (k_{B} \beta)^{-1}$ where $k_B$ is the Boltzmann constant henceforth set to $1$ and chemical potential $\mu$, 
\begin{equation}
\label{eq:rhoth}
\rho^{\rm th}_{B}=\frac{e^{-\beta(H_B - \mu N_{B})}}{Z}
\end{equation}
where $N_B$ is the number operator for the bath and $Z$ is the grand partition function for the bath. The superscript ``th'' in Eq.~\eqref{eq:globalrho} and Eq.~\eqref{eq:rhoth} stands for thermal equilibrium. 

With this generic setup in hand, our interest here is to study the quantum dynamics of spatial density profile of bosons/fermions
\begin{equation}
\label{eq:spatial}
    n_{i} (t) = \big\langle {a}^{\dagger}_{i}(t) {a}_{i}(t) \big\rangle,
    \end{equation} 
where we use the Heisenberg representation for the operators and $\langle...\rangle$ denotes an average over the initial density matrix $\rho(0)$. We also look at the total number of particles in the lattice  
\begin{equation}
\label{eq:totalN}
    N(t) = \sum_{i = 1}^{L} n_{i} (t).
\end{equation}

In order to have a detailed understanding of the quantities given in Eq.~(\ref{eq:spatial}) and Eq.~(\ref{eq:totalN}), we use  four different approaches which we elaborate below. 

\subsection{Method 1: Exact quantum dynamics for correlation matrix}
\label{M1}
In this subsection, we discuss direct exact numerical calculation for computing the observables of interest. As the entire setup here is quadratic, the equations of motion for the two-point correlation functions involving both the system and bath degrees of freedom closes. Thus, the central idea here is to numerically evolve the two-point correlation function consisting of all the degrees of freedom via the single particle Hamiltonian of the setup. Once this unitary propagation is performed, the quantities mentioned in  Eq.~(\ref{eq:spatial}) and Eq.~(\ref{eq:totalN}) can be suitably extracted. This procedure, of course, involves one to consider a large but finite bath. 

Let $L_{B}$ be the number of bath sites such that $L_{B} \gg L$. We write the full Hamiltonian for the setup as
\begin{equation}
\label{eq: fullH}
H = H_{S} + H_{B} +   {H}_{SB} = \sum_{i,j=1}^{L+L_B} h_{i j}  \, {d}^{\dagger}_{i} \,  {d}_{j}\equiv D^{\dagger}\, h \,D.
\end{equation}
Here, $D = \{  {d}_{i}\}$ is a column vector containing all the annihilation operators of the system and the bath. Specifically, $D \equiv \{  {a}_{1},   {a}_{2}, ...,   {a}_{L},   {b}_{1},   {b}_{2}, ...,   {b}_{L_{B}}\}$. Naturally, $D^{\dagger} = \{  {d}^{\dagger}_{i}\}$ is the row vector consisting of all the creation operators of the system and the bath. $h$ is the single particle Hamiltonian of the full setup and has dimension $(L+L_B) \times (L+L_B)$. Since our central focus here is to study the filling of lattice system, we will not require the information of the full density matrix. Instead, we will just need to focus on the correlation matrix 
\begin{equation}
\label{eq:S}
    S(t) = \Big\langle \big[D^{\dagger}(t)\big]^T \big[D^{T}(t)\big] \Big\rangle
\end{equation}
where the superscript $T$ stands for the transpose of a matrix. The matrix element of $S$ is given as 
\begin{equation}
\label{eq:Sij}
    S_{ij} (t) = \langle   {d}^{\dagger}_{i}(t) {d}_{j}(t)\rangle.
\end{equation}
Following  the Heisenberg equation of motion $i \, \frac{d}{dt} {O} = \big[O, H\big]$, for any operator $O(t)$,  one can write 
\begin{equation}
\label{eq:commu}
    \frac{d}{dt}\Big({d}^{\dagger}_{i} {d}_{j}\Big) = i \Big[H,  {d}^{\dagger}_{i} {d}_{j}\Big]
   = i \sum_{r,s=1}^{L+L_B} h_{rs} \Big[{d}^{\dagger}_{r} {d}_{s}, {d}^{\dagger}_{i} {d}_{j}\Big].
\end{equation}
Using the commutation and anti-commutation relations for bosons and  fermions respectively, Eq.~\eqref{eq:commu} can be simplified to 
\begin{equation}
    \frac{d}{dt}\Big({d}^{\dagger}_{i} {d}_{j}\Big) =
         i \sum_{r=1}^{L+L_B} \Big(h_{r i } {d}_{r}^{\dagger}{d}_{j} - h_{j r} {d}^{\dagger}_{i} {d}_{r} \Big).
    \label{eq:two-op}
\end{equation}    
Eq.~\eqref{eq:two-op} holds for both bosons and fermions. Using Eq.~(\ref{eq:two-op}), and the fact that the single particle Hamiltonian $h$ is symmetric, we can obtain the evolution for the correlation matrix element of $S(t)$, which is given by \cite{Auditya_exact,Junaid_transport,purkayastha2016out} 
\begin{equation}
    \dot{S}_{ij} (t) = i \sum_{r=1}^{L+L_{B}} \Big(h_{i r} \,S_{r j}(t) - S_{i r} (t)\, h_{r j} \Big) = i \Big[h, S(t)\Big]_{ij}
\end{equation}
the solution of which is given by,
\begin{equation}
\label{eq:Sprop}
S(t) = e^{i h t} S(0) e^{-i h t}   
\end{equation}
where we recall that $h$ is the single particle Hamiltonian for the full setup. Eq.~\eqref{eq:Sprop} holds for both bosons and fermions. We can easily construct the initial correlation matrix $S(0)$ from the initial density operator $\rho(0)$. Since the lattice chain is initially in vacuum and is decoupled from the bath, all the two-point correlations of the form $\langle   {a}_{i}^{\dagger}(0)  {a}_{j}(0) \rangle$, $\langle   {a}_{i}^{\dagger}(0)  {b}_{j}(0) \rangle$ and $\langle   {b}_{i}^{\dagger}(0)   {a}_{j}(0) \rangle$ will be zero. The non-zero entries in $S(0)$ are the two-point correlations of the bath degrees of freedom and of the form $\langle   {b}_{i}^{\dagger}(0)  {b}_{j}(0) \rangle$. These entries can be obtained as follows. We recall that the bath Hamiltonian given in Eq.~(\ref{eq:bath}) can be expressed as, 
\begin{equation}
    H_{B} = \sum_{i,j=1}^{L_{B}} h^{B}_{ij}   {b}^{\dagger}_{i}   {b}_{j} 
\end{equation}
where $h^{B}$ is the single particle Hamiltonian of the bath. This Hamiltonian can be easily diagonalized by a unitary transformation $U$, i.e., $h^{B} = U \Lambda_{B} U^{\dagger}$ with $\Lambda_{B}$ being a diagonal matrix with single particle eigenvalues as its entries. The Hamiltonian in the diagonal form can be written as 
\begin{equation}
    H_B=  \sum_{q=1}^{L_{B}} \lambda^{B}_{q} \,  {b}_{q}^{\dagger} \, {b}_{q}
    \end{equation}
where 
\begin{equation}
{b}_{i} = \sum_{q=1}^{L_{B}} U_{i q} \, {b}_{q}
\end{equation}
with ${b}_{q}$ being the annihilation operator of $q$-{th} normal mode of the bath with eigenvalues $\lambda^{B}_{q}$. One can then easily find that
\begin{equation}
\label{LIQP4}
    \langle   {b}^{\dagger}_{i}(0)   {b}_{j}(0) \rangle^{\rm th} = \sum_{q=1}^{L_{B}} U^{*}_{i q}\, U_{j q} \, \bar{n}(\lambda^{B}_{q}),
\end{equation}
where we have used the fact that $\langle   {b}^{\dagger}_{q}   {b}_{q'} \rangle = \bar{n}(\lambda^{B}_{q}) \, \delta_{q q'}$. Here $\bar{n}(\omega)$ can either be Bose or  Fermi function and is given by 
\begin{equation}
\label{eq:nw}
    \bar{n}(\omega) = \frac{1}{e^{\beta (w-\mu)} \pm 1} 
\end{equation}
where $-$ and $+$ stands for bosons and fermions, respectively. With $S(t=0)$ constructed from all these initial correlations, we can now propagate the correlation matrix following Eq.~(\ref{eq:Sprop}) and suitably extract the required entries from $S(t)$ to compute $n_i(t)$ and thereby $N(t)$, as defined in Eq.~(\ref{eq:spatial}) and Eq.~(\ref{eq:totalN}), respectively. It should be noted that if $L_{B} \gg L$, then the system dynamics is almost equivalent to that when subjected to a true bath with infinite degrees of freedom. 

Since this exact numerical recipe involves unitary evolution with respect to the Hamiltonian of the entire setup [see Eq.~(\ref{eq: fullH})], it can become computationally difficult if the total size $L+L_{B}$ is large. Therefore, it is useful to study time dynamics for this setup following complementary approaches, namely the Redfield and Lindblad master equations which are perturbative and Markovian in nature. This procedure involves integrating out  infinite degrees of freedom of the bath and thereby providing an effective dynamical description for the reduced density matrix of the lattice system which can then be used to study lattices with large number of sites. 

\subsection{Method 2: Redfield Quantum Master Equation Approach}
\label{M2}
In this subsection, we discuss the Redfield equation and provide the key steps that are involved to obtain the spatial density profile and total number of particles. We start by writing the system-bath interaction Hamiltonian, given in Eq.~(\ref{LIQP2}), in the interaction picture as,
\begin{eqnarray}
 H^{I}_{SB}(t) &=& e^{i H_{0} t}   {H}_{SB} e^{-i H_{0} t} \nonumber \\
 &=&  \gamma \left(  {a}^{\dagger}_{m} (t)   {b}_{1} (t) +  {b}^{\dagger}_{1} (t) {a}_{m}(t)    \right)
\end{eqnarray}
where $H_0=H_S+H_B$ and
\begin{equation}
a_{m} (t) = e^{i H_{S} t}   {a}_{m} e^{-i H_{S} t}; \quad  b_{1} (t) = e^{i H_{B} t}   {b}_{1} e^{-i H_{B} t}.  
\end{equation}
Starting from the exact von-Neumann equation, one can write an exact equation governing the dynamics of the reduced density matrix for the system in the interaction picture as
\begin{equation}
\label{NP6} 
\frac{d}{dt} \rho_{SI} (t) = - \int_{0}^{t} d\tau \operatorname{Tr_{B}}\Big[H^{I}_{SB} (t), \big[H^{I}_{SB} (\tau), \rho_{I} (\tau)\big]\Big],
\end{equation}
where $\rho_I(\tau)$ is the full density operator in the interaction picture.  The subscript ``$SI$" in Eq.~\eqref{NP6} stands for system and interaction picture, respectively. Now to arrive at the Redfield equation, one assumes (i) weak system-bath coupling limit (Born approximation) and (ii) Markovian  limit \cite{breuer2002theory, carmichael2009statistical,agarwal2012quantum}. The Born approximation implies writing $\rho_{I} (\tau)$ in Eq.~(\ref{NP6}) as a direct product state of the system and the bath density matrix i.e., $\rho_{I} (\tau)= \rho_{SI} (\tau) \otimes \rho^{\rm th}_{B}$ where $\rho^{\rm th}_{B}$ is defined in Eq.~(\ref{eq:rhoth}). 
The Markov approximation involves changing $\rho_{SI} (\tau)$ to $\rho_{SI} (t)$ and further extending the upper limit of the integral $t$ to $\infty$, in Eq.~(\ref{NP6}). After some algebraic manipulations, we obtain the Redfield equation as \cite{purkayastha2016out} 
\begin{widetext}
\begin{equation}
\label{TBPI5}
\begin{aligned}
    \frac{d}{dt} \rho_{SI} (t) &= - \gamma^{2} \int_{0}^{\infty} d\tau \left[\Big\langle b_{1} (t) b^{\dagger}_{1} (\tau)\Big \rangle \Big[a^{\dagger}_{m} (t), a_{m} (\tau) \rho_{SI}(t)\Big] + \Big\langle b^{\dagger}_{1} (\tau) b_{1} (t) \Big\rangle \Big[\rho_{SI}(t) a_{m} (\tau), a^{\dagger}_{m}(t)\Big] + h.c. \right].
\end{aligned}
\end{equation}
In the Schr\"odinger picture, we receive,
\begin{equation}
\label{LIQP5}
\begin{aligned}
    \frac{d}{dt} \rho_{SS} (t)  &= i\Big[\rho_{SS}, H_{S}\Big] \\
    & \quad - \gamma^{2} \int_{0}^{\infty} d\tau \left[\langle b_{1} (t) b^{\dagger}_{1} (\tau) \rangle \Big[  {a}^{\dagger}_{m}, a_{m} (\tau - t) \rho_{SS}(t)\Big] + \langle b^{\dagger}_{1} (\tau) b_{1} (t) \rangle \Big[\rho_{SS}(t) a_{m} (\tau - t), {a}^{\dagger}_{m}\Big] + h.c. \right],
\end{aligned}
\end{equation}
\end{widetext}
where the subscript ``$SS$" in Eq.~\eqref{LIQP5} stands for system and Schr\"odinger picture, respectively. Since the bath operators are defined in the interaction picture, the corresponding two-point correlation functions are known exactly and as before [see Eq.~\eqref{LIQP4}] can be written in terms of the normal modes of the bath as,
\begin{equation}
\label{TBPI7}
\begin{aligned}
    \langle   {b}^{\dagger}_{1} (\tau)   {b}_{1} (t)\rangle &= \sum_{q} |U_{1 q}|^{2}\, e^{- i \lambda^{B}_{q} (t-\tau)}\, \bar{n}(\lambda^{B}_{q}),\\
    \langle   {b}_{1} (t)   {b}^{\dagger}_{1} (\tau) \rangle &= \sum_{q} |U_{1 q}|^{2}\, e^{- i \lambda^{B}_{q} (t-\tau)}\, \big[1 \pm \bar{n}(\lambda^{B}_{q})\big],
\end{aligned}
\end{equation}
where recall that, $\lambda_q^B$ is the eigenvalue of the $q$-th mode of the bath and $\bar{n}(\omega)$ is defined in Eq.~(\ref{eq:nw}). In Eq.~\eqref{TBPI7}, the $\pm$ stands for bosons and fermions, respectively. We  now express the above Redfield equation in Eq.~(\ref{LIQP5}) in terms of the eigenoperators of the system Hamiltonian. In other words, we first diagnolize the lattice Hamiltonian $H_{S}$ and write,
\begin{equation}
\begin{aligned}
H_{S} &= \sum_{i, j =1}^{L} h^{S}_{ij}   {a}_{i}^{\dagger}   {a}_{j} = \sum_{k=1}^{L} \lambda^{S}_{k}   {a}_{k}^{\dagger}   {a}_{k}
\end{aligned}
\end{equation}
where 
\begin{equation}
    h^{S} = W \Lambda_{S} W^{\dagger}
\end{equation} 
and therefore the matrix $W$ is responsible for diagonalizing the single-particle system Hamiltonian $H_S$ and $\Lambda_{S}$ is the diagonal matrix containing the single particle eigenvalues of the system. Here 
\begin{equation}
\label{eq:a_i_k}
{a}_{i} = \sum_{k=1}^{L} W_{i k} {a}_{k}.
\end{equation} 
Following this diagonalization procedure and using Eq.~(\ref{LIQP5}) and  Eq.~(\ref{TBPI7}), we can rewrite the Redfield equation as \cite{purkayastha2016out},
\begin{equation}
\label{TBPI9}
\dot{\rho}_{SS} (t) \!=\! i\Big[\rho_{SS}, H_{S}\Big] \!-\! \sum_{k, k'=1}^{L} \int_{-\infty}^{\infty} \frac{d \omega}{2 \pi} \Big[I(\omega) \mathcal{L} (\rho_{SS}) + h.c.\Big]
\end{equation}
where 
\begin{equation}
\label{eq:Iw}
 I(\omega) = \int_{0}^{\infty} \, d\tau \, e^{- i (\omega - \lambda^{S}_{k'}) \tau},
\end{equation}
and 
\begin{eqnarray}
\label{eq:L}
    \mathcal{L} (\rho_{SS}) &= \Big[f_{k k'} (\omega) \pm F_{k k'} (\omega)\Big] \Big[  {a}^{\dagger}_{k},   {a}_{k'} \rho_{SS}(t)\Big]\nonumber \\
    &  + F_{kk'} (\omega) \Big[\rho_{SS}(t)   {a}_{k'},   {a}^{\dagger}_{k}\Big].
\end{eqnarray}
The functions $f_{kk'}(\omega)$ and $F_{kk'}(\omega)$ in Eq.~(\ref{eq:L}) are defined as, 
\begin{eqnarray}
    f_{k k'} (\omega) &=& W^{*}_{m k} \,W_{m k'} \,J(\omega),\nonumber \\ 
    F_{k k'} (\omega) &=& W^{*}_{m k}\, W_{m k'} \,J(\omega)\, \bar{n}(\omega),
\end{eqnarray}
where recall that the index $m$ in $W_{m k}$ refers to the $m$-th site of the lattice system that is connected with the bath [see Eq.~(\ref{LIQP2})]. Note that $\pm$ sign in Eq.~(\ref{eq:L}) again refers to the boson/fermion case. Here, $J(\omega)$ is the spectral density of the bath, defined as, 
\begin{equation}
\label{eq:spectral}
    J(\omega) \equiv 2 \pi \, \gamma^{2} \sum_{q} |U_{1 q}|^{2} \delta(\omega - \lambda^{B}_{q}).
\end{equation}
Note that two-point correlation functions of the system are given as
\begin{equation}
    \label{eq:Rcor}
    C_{k,k'} (t) = \langle   {a}^{\dagger}_{k}(t)   {a}_{k'}(t) \rangle = {\rm Tr} \big[{a}^{\dagger}_{k}(0)   {a}_{k'}(0) \rho_{SS}(t)\big].
\end{equation}
From the Redfield equation [Eq.~(\ref{TBPI9})], one can write down a differential equation for the two-point correlation function defined in Eq.~(\ref{eq:Rcor}) as \cite{MRF-OQS,PhysRevA.82.013640,wu2010non},
\begin{eqnarray}
\label{TBPI11}
    \frac{d C_{k,k'}(t)}{dt} = i \lambda_{k}^{S} C_{k,k'} (t) &+& \frac{1}{2} \left[\tilde{F}_{k'k} - \sum_{\bar{k}=1}^{L} \tilde{f}_{k' \bar{k}} \, C_{k, \bar{k}} (t) \right]\nonumber \\  \qquad &+& (k \Longleftrightarrow k')^{\dagger},
\end{eqnarray} 
where $(k \Longleftrightarrow k')^{\dagger}$ is a short form for the right hand side of Eq.~\eqref{TBPI11} when $k$ and $k'$ are interchanged and the terms are subjected to complex conjugation.  The new functions (denoted by the symbol tilde) in Eq.~\eqref{TBPI11} are defined as,
\begin{eqnarray}
\label{eq:f}
    \tilde{f}_{k' \bar{k}} &= f_{k' \bar{k}} (\lambda_{\bar{k}}^{S}) - i \, P \int_{-\infty}^{\infty} \frac{d \omega}{\pi} \frac{f_{k' \bar{k}} (\omega)}{\omega - \lambda_{\bar{k}}^{S}},\\ \tilde{F}_{k' k} &= F_{k' k} (\lambda_{k}^{S}) - i \, P \int_{-\infty}^{\infty} \frac{d \omega}{\pi} \frac{F_{k' k} (\omega)}{\omega - \lambda_{k}^{S}}.
    \label{eq:F}
\end{eqnarray}
Here $P$ refers to the Cauchy principle value. This particular form in Eq.~(\ref{eq:f}) and Eq.~(\ref{eq:F}) is obtained by writing $I(\omega)$ in Eq.~(\ref{eq:Iw}) using the Sokhotski-Plemelj theorem. Eq.~(\ref{TBPI11}) forms a closed set of differential equations for the two-point correlation function which can be solved numerically by grouping the equations in a matrix equation form. We therefore write the components of $C_{k, k'} (t)$ as a column vector with elements  ${\bf C}_r$, $r = 1,2,\ldots,L^2$ and denote it by $\vec{\bf C}(t)$, and write Eq.~(\ref{TBPI11}) as
\begin{equation}
\label{LIQP6}
    \frac{d \vec{\bf C}}{dt} = M \vec{\bf C} + \vec{\bf Q}
\end{equation}
where $M$ is the homogeneous part and is a $L^2 \times L^2$ matrix and $\vec{\bf{Q}}$ is a $L^2 \times 1$ column vector. Note that the information about the quantum statistics is encoded only in the column vector $\vec{\bf{Q}}$ as a consequence of which the quantum dynamics of fermions and bosons differ.
The formal solution to Eq.~(\ref{LIQP6}) with the initial condition $\vec{\bf{C}}(0)=0$ (note that the lattice is initially empty) is given by,
\begin{equation}
\label{eq:vecC}
    \vec{\bf C}(t)=\int_{0}^{t} \, d\tau \,e^{M(t-\tau)} \vec{\bf Q}.
\end{equation}
We now write the solution in Eq.~(\ref{eq:vecC}) more explicitly. To do so, we first diagonalise $M$ as $M =V \Lambda_{M} V^{-1}$, and obtain
\begin{equation}
\begin{aligned}
   {\bf C}_{r} (t) &= \int_{0}^{t} d\tau \sum_{\alpha,s=1}^{L^2} V_{r \alpha} e^{\lambda^{M}_{\alpha} (t-\tau)} {V^{-1}_{\alpha s}} {\bf Q}_{s}\\ &= \sum_{\alpha,s=1}^{L^2}\,\Bigg(\frac{e^{\lambda^{M}_{\alpha} t} - 1}{\lambda^{M}_{\alpha}}\Bigg)\, V_{r \alpha} {V^{-1}_{\alpha s }} \, {\bf Q}_s. 
\end{aligned}
\label{eq:corr-eig}
\end{equation}
 The real part of the eigenvalues $\{\lambda_\alpha^M\}$ of $M$ matrix are expected to be all negative which would ensure a well-defined steady state in the long-time limit. We now study the short and long-time limits of ${\bf C}_{r} (t)$. 
 Let us denote the eigenvalue with the largest magnitude as $\lambda_L$ and the one with the smallest real part magnitude as $\lambda_S$. In the short time limit $(t \ll 1/|\lambda_L|)$, we can do a Taylor expansion and find that all  two-point correlations  grow linearly with time, thus
\begin{equation}
\label{LIQP10}
   {\bf C}_{r} (t \ll 1) 
   = \chi_{r} \, t +O(t^2),
    \end{equation}
with a slope
\begin{equation}
\label{eq:chi}
\chi_r = \Bigg(\sum_{\alpha,s=1}^{L^2} V_{r \alpha} {V^{-1}_{\alpha s}} {\bf Q}_{s}\Bigg)={\bf Q}_r.
\end{equation}
We now discuss the long-time limit i.e., $t \to \infty$. Note that Eq.~(\ref{eq:corr-eig}) can be recasted as,
\begin{equation} 
   {\bf C}_{r} = {\bf C}^{SS}_r + \sum_{\alpha,s=1}^{L^2} V_{r \alpha} {V^{-1}_{\alpha s}} {\bf Q}_{s} \frac{e^{\lambda^{M}_{\alpha} t}}{\lambda^{M}_{\alpha}},
   \label{eq:c-recast}
\end{equation}
where the steady state correlation elements ($t \to \infty$) are given by,
\begin{equation}
    {\bf C}_r^{SS} = -\sum_{\alpha,s=1}^{L^2} \frac{V_{r \alpha} {V^{-1}_{\alpha s }} {\bf Q}_{s}}{\lambda^{M}_{\alpha}} = -(M^{-1} \vec{\bf Q})_r,
\end{equation}
and the second term in Eq.~(\ref{eq:c-recast}) indicates a long time  exponential approach to the steady state. The  eigenvalue, $\lambda_S$,  with the smallest magnitude for the real part  will determine the time scale, $1/|{\rm Re}[\lambda_S]|$, for convergence to the steady state. As, the correlation functions are obtained in the diagonalized basis, to determine the spatial density profile, the final step is to come back to the local site basis which gives, 
\begin{equation}
\label{eq:n_i_rotated}
    n_i(t) = \langle {a}^{\dagger}_{i}(t) {a}_{i}(t) \rangle = \sum_{k,k'=1}^{L} W_{i k'} W^{*}_{i k} C_{k,k'}(t)
\end{equation}
where recall that $C_{k,k'}$ is defined in Eq.~(\ref{eq:Rcor}).
The total particle number is given by summing over all lattice sites, 
\begin{equation}
N(t)=\sum_{k=1}^{L} C_{k,k}(t)
\end{equation}
which at early times $(t \ll 1)$ gives
\begin{align}
\label{eq:early-time}
N(t) \approx t   \sum_{k=1}^{L} Q_{k,k}&= t \sum_{k=1}^{L} \tilde{F}_{k,k}\\&=  t \sum_{k=1}^{L} |W_{m k}|^2 \, J(\lambda_k^S)\, \bar{n}(\lambda_k^S).
\end{align}
Eq.~\eqref{eq:early-time} clearly demonstrates an early time linear growth with different slopes for fermions and bosons. In the limiting case with very small inter-site hopping $g$, one can set $\lambda_k^S\approx 0$ (the eigenvalues of uncoupled lattice sites), as a result of which we get
\begin{align}
N(t) \approx t \,J(0) \, \bar{n}(0).
\end{align}
We will later see that this is exactly what one receives from the local Lindblad equation. Note that in a suitable parameter regime, the Redfield approach can be simplified to a local Lindblad equation. As we will show in the next subsection [Sec.~\ref{M3}], this allows for elegant analytical expressions for the local density $n_i(t)$ in Eq.~\eqref{eq:spatial} and the total occupation $N(t)$ in Eq.~\eqref{eq:totalN}.

\subsection{Method 3: Lindblad approach}
\label{M4}
In this subsection, we outline the Lindblad approach and present our results for local density $n_i(t)$ in Eq.~\eqref{eq:spatial} and total occupation $N(t)$ in Eq.~\eqref{eq:totalN}. A common way to model open quantum systems that mimics incoherent processes is via the local Lindblad formalism \cite{lindblad1976generators, gorini1976completely, PhysRevA.105.032208,purkayastha2016out,manzano2020short} which is of the form 
 \begin{equation}
\label{eq:Lind}
    \dot{\rho}_{SS} (t) = i\Big[\rho_{SS}, H_{S}\Big] + \mathcal{D}\big[\rho_{SS} (t)\big]
    \end{equation}
    where ${\mathcal{D}}$ is the Lindbladian and is given by,
    \begin{equation}
    \label{eq:dissi}
    \mathcal{D}\big[\rho_{SS} (t)\big] = 2\, {\cal O} \rho_{SS} {\cal O}^{\dagger}- \big\{{\cal O}^{\dagger} {\cal O}, \rho_{SS}\big\},
\end{equation}
where we recall that $\rho_{SS}(t)$ is the reduced system density matrix in the Schr\"odinger picture and $H_S$ is the system Hamiltonian, given in Eq.~(\ref{eq:hs}). 
Here  ${\cal O}$ represents different channels of openness of the lattice system. For our setup, if we derive a local Lindblad equation starting from the fully microscopic system-reservoir Hamiltonian [Eq.(\ref{eq:hs}), Eq.~(\ref{eq:bath}), Eq.~(\ref{LIQP2})], both incoherent pump and loss terms naturally arise in the Lindbladian given in Eq.~(\ref{eq:Lind}). More explicitly, the systematically derived local Lindblad equation \cite{lindblad1976generators, gorini1976completely, PhysRevA.105.032208,purkayastha2016out,manzano2020short} is given as 
\begin{eqnarray}
\label{eq:mod-Lind}
    \dot{\rho}_{SS} &= i \Big[\rho_{SS}, H_{S}\Big] + \Gamma_G \Big[2 {a}^{\dagger}_{m} \rho_{SS} {a}_{m}\! - \! \{{a}_{m} {a}_{m}^{\dagger}, \rho_{SS} \} \Big]\nonumber \\ & + \Gamma_L \Big[2 {a}_{m} \rho_{SS} {a}_{m}^{\dagger} - \{{a}_{m}^{\dagger} {a}_{m}, \rho_{SS} \}\Big]
\end{eqnarray}
where the gain $\Gamma_G$ and the loss $\Gamma_L$ coefficients are given by
\begin{eqnarray}
\label{eq:GammaG}
    \Gamma_G &=& \frac{J(0)}{2} \, \bar{n}(0),\\
    \Gamma_L &=&  \frac{J(0)}{2} \, \big(1 \pm \bar{n}(0)\big).
    \label{eq:GammaL}
\end{eqnarray}
Recall that $J(\omega)$ is the spectral density of the bath, defined in Eq.~\eqref{eq:spectral}. Note that the zeros in the argument of $J(0)$ and $\bar{n}(0)$ in Eq.~(\ref{eq:GammaG}) and Eq.~(\ref{eq:GammaL}) are due to the fact that there is no onsite term in the system Hamitonian given in Eq.~(\ref{eq:hs}). The $\pm$ sign in Eq.~(\ref{eq:GammaL}) stands for bosons and fermions, respectively. It is important to highlight that the validity of local Lindblad equation in Eq.~\eqref{eq:mod-Lind} relies on weak system bath coupling $\gamma$ as well as weak inter-site hopping parameter $g$ within the lattice system \cite{purkayastha2016out}.

Following Eq.~\eqref{eq:mod-Lind} one can write down the equations of motion for the two-point correlation functions of the system which is defined as 
\begin{equation}
\label{eq:correlation}
C_{i,j}= \langle a_i^{\dagger} a_j \rangle.
\end{equation}
The equations of motion is given by,
\begin{eqnarray}
\label{eq:mod-corr-Lind}
    \frac{d C_{i,j}}{dt} &= i \, g\, (C_{i-1, j} \!-\! C_{i, j+1} \!+\! C_{i+1, j} \!-\! C_{i, j-1})\nonumber \\ 
    & -(\Gamma_L \mp \Gamma_G) (\delta_{i m} + \delta_{j m})\, C_{i, j} + 2\, \Gamma_G \delta_{m i} \delta_{m j} \nonumber \\
\end{eqnarray}
where $\mp$ stands for bosons and fermions, respectively. It is crucial to note that $\Gamma_G$ [Eq.~(\ref{eq:GammaG})] and $\Gamma_L$ [Eq.~(\ref{eq:GammaL})] here are related by detailed balance i.e., they are not independent of each other.

We now solve for correlation functions in Eq.~\eqref{eq:mod-corr-Lind} and subsequently extract local population and total occupation. Without loss of generality, in this subsection, we consider the bath to be attached to the lattice at site $m=0$. Furthermore, we take the lattice size $L$ to be infinity, i.e., the lattice is now extended from $-\infty$ to $+\infty$. Since the system is expected to be symmetric about the $0$-th site, for simplicity, we consider only the positive side of the lattice chain in the analysis presented below. 
The spatial density profile $n_i(t)$ is given by [see Appendix \ref{app:scaling} for the details],
\begin{equation}
\label{eq:ni_int_exact-m}
    n_i(t) = 2\, \Gamma_G \, \int_{0}^{ t} d\tau \, |\tilde{S}_i(\tau)|^2
    \end{equation}
where for large $\tau$,
\begin{equation}
\label{eq:scaledS-m}
    \tilde{S}_i(\tau) = \frac{i\, J_i(2 \, g \, \tau)}{i + \, \tau \Gamma^{'}}.
\end{equation}
Here $J_i$ is the Bessel function of first kind and 
\begin{equation}
\label{eq:gprime_main}
    \Gamma^{\prime} = \Gamma_L \mp \Gamma_G = \frac{J(0)}{2}
\end{equation}
where $\mp$ stands for bosons and fermions, respectively and we recall that $\Gamma_G$ and $\Gamma_L$ are given in Eq.~(\ref{eq:GammaG}) and Eq.~(\ref{eq:GammaL}), respectively.  
Interestingly, $\Gamma^{'}$ defined in Eq.~\eqref{eq:gprime_main} is independent of the statistics and is always positive.  As a consequence, the quantum statistics is encoded as a prefactor ($\Gamma_G$) in the density profile. 

Interestingly, in the limit $i \to \infty$, $t\to \infty$ while keeping ${i}/{t}$ as fixed, the analytical results in Eqs.~\eqref{eq:ni_int_exact-m} and  \eqref{eq:scaledS-m}  admit an interesting scaling form  [see Appendix.~(\ref{app:scaling}) for details]
\begin{equation}
\label{eq:ni_scale-main-1}
     n_i(t) = \Phi\left(\frac{i}{2 g t}\right)
\end{equation}
where the scaling function $\Phi(\nu)$ is exactly given by 
\begin{equation}
\label{eq:ni_scale-main-2}
    \Phi(\nu) =  \frac{4\,\Gamma_G \,g}{\pi}  \,\int_{1}^{\frac{1}{\nu}} dz\, \frac{1}{\sqrt{z^2-1}} \, \frac{1}{(2\, g + z\,\Gamma^{'})^2}.
\end{equation}  
The integral in Eq.~\eqref{eq:ni_scale-main-2} can be performed exactly to yield 
\begin{widetext}
\begin{equation}
\label{eq:phi-scale-int}
\Phi(\nu) = \frac{{\tilde{g}} \, ( 1 + \nu \, \tilde{g}) \Big[\log (1 +  \nu \, \tilde{g} )-\log \left(\tilde{g}+\nu -\sqrt{\left(\tilde{g}^2-1\right) \left(1-\nu ^2\right)} \right)\Big]-\sqrt{\left(\tilde{g}^2-1\right) \left(1-\nu ^2\right)}}{\left(\tilde{g}^2-1\right)^{3/2} (\nu\, \tilde{g} +1)}, \quad 0 < \nu < 1.
\end{equation}
\end{widetext}
where the dimensionless parameters $\tilde{g}$ is given by 
\begin{equation}
    \label{eq:gtilde}
\tilde{g}=\frac{2g}{\Gamma'} = \frac{4g}{J(0)}
\end{equation}
and we recall that $\Gamma'$ is given by Eq.~\eqref{eq:gprime_main} and $J(0)$ from Eq.~\eqref{eq:jw_sqrt} is given by
\begin{equation}
J(0) = \frac{2 \gamma^2}{t_B}.
\end{equation} 
Note that the scaling function $\Phi(\nu)$ in Eq.~\eqref{eq:phi-scale-int} admits the following limiting forms near $\nu \to 0$ and $\nu \to 1$ :
\begin{equation}
\Phi(\nu) = \frac{1}{1-\tilde{g}^2}\!-\!\frac{\tilde{g} \log
   \left(\tilde{g}\!-\!\sqrt{\tilde{g}^2-1}\right)}{\left(\tilde{g}^2-1\right)^{3/2}} \!-\! \frac{\nu^2}{2} + O(\nu^3) 
\end{equation}
when $\nu \ll 1$. Therefore the decay of $\Phi(\nu)$ from the peak at $\nu=0$ is parabolic in nature. On the other hand, for $\nu \to 1$ we get, 
\begin{equation}
\label{eq:nu1density}
  \Phi(\nu) = \frac{\sqrt{2 (1-\nu)}}{1+ \tilde{g}^2} + O\big[(1-\nu)^{3/2}\big]\quad\text{as}~\nu\to 1.
 \end{equation}
From Eq.~\eqref{eq:nu1density}, it is interesting to note that the scaled density $\Phi(\nu)$ vanishes in a square root form.  It is worth noting that the analytical scaling form in Eq.~\eqref{eq:ni_scale-main-2} is independent of quantum statistics except from the prefactor ($\Gamma_G$).

The total particle number $N(t)$ is given by [see Appendix.~(\ref{app:scaling}) for details] 
\begin{widetext}
\begin{align}
N(t)&=  \frac{4\,\Gamma_G \,\tilde{g}^2 t }{\pi}  \,\int_{1}^{\infty} \frac{dz}{z}\, \frac{1}{\sqrt{z^2-1}} \, \, \frac{1}{(\tilde{g} + z)^2} 
=-\frac{2\,\Gamma_G \,t }{\pi (1-\tilde{g}^2)} \,  \left[2 \tilde{g}-\pi(1-\tilde{g}^2) + 2\big(1-2 \tilde{g}^2\big) \frac{\cos^{-1}\big(\tilde{g}\big)}{\sqrt{1-\tilde{g}^2}}\right].
\label{eq:N_t_scaling-m}
\end{align}
\end{widetext}
Note that using the relation 
\begin{equation}
    \label{eq:cos}
\frac{\cos^{-1}\big(\tilde{g}\big)}{\sqrt{1-\tilde{g}^2}} = - \frac{\log\big(\tilde{g} +\sqrt{\tilde{g}^2-1} \big)}{\sqrt{\tilde{g}^2-1}}
\end{equation}
it is easy to see that Eq.~\eqref{eq:N_t_scaling-m} is always real for all values of $\tilde{g}$. From Eq.~\eqref{eq:N_t_scaling-m} it is clear that $N(t)$ always exhibits a linear growth in time. One can further simplify  Eq.~\eqref{eq:N_t_scaling-m} in the limit of small and large $\tilde{g}$. We get
\begin{equation}
\label{eq:N_g_small}
    N(t) = t \, \Gamma_G\, \Big[\tilde{g}^2 - \frac{16}{3 \pi} \tilde{g}^3 \Big] + O(\tilde{g}^4) 
\end{equation}
when $\tilde{g} \ll 1$ and 
\begin{equation}
\label{eq:N_g_large}
    N(t) = t \, \Gamma_G\, \Big[2 + \frac{4}{\pi} \,\frac{1 \!-\! \log(4) \!-\! 2 \log (\tilde{g}) \big]}{\tilde{g}} \Big] + O\Big(\frac{1}{\tilde{g}^3}\Big), 
\end{equation}
when $\tilde{g} \gg 1$.

Note that although Redfield [Sec.~\ref{M2}] and Lindblad [Sec.~\ref{M4}] equations offers us the advantage of tracking both time dynamics and steady state, it involves perturbative and Markovian approximations. To get an analytical handle of the steady state, via fully non-perturbative approach, we now discuss the exact steady state using  quantum Langevin approach.

\subsection{Method 4: Quantum Langevin Equation approach}
\label{M3}
In this subsection, we discuss the quantum Langevin equation (QLE) approach \cite{dhar2006nonequilibrium,dhar2012nonequilibrium,dhar2006heat,Wang_QLE, Ford_QLE,segalQLE, QLE_Talkner, QLE_Lebowitz, bondyopadhaya2022nonequilibrium}. Given the bilinear nature of the entire setup, we can compute exactly the steady state properties of the lattice chain following this approach. 
Let us start by re-writing the Hamiltonian in Eqs.~(\ref{eq:hs}, \ref{eq:bath}, \ref{LIQP2}) as,
\begin{eqnarray}
H_{S} &=& \sum_{i,j = 1}^{L} h^{S}_{i j } {a}^{\dagger}_{i} {a}_{j}, \quad H_{B} = \sum_{i,j = 1}^{L_{B}} h^{B}_{ij} {b}^{\dagger}_{i} {b}_{j}, \nonumber \\
{H}_{SB} &=& \sum_{i=1 }^{L}\sum_{j = 1}^{L_{B}} h^{SB}_{ij} {a}^{\dagger}_{i} {b}_{j} + {h.c.}
\end{eqnarray}
Let us denote $A(t) = \{a_{i} (t) \}$ and $B(t) = \{b_{i} (t) \}$ as the column vectors consisting of system and bath annihilation operators, respectively. The Heisenberg equation of motion for the respective components are given as:
\begin{eqnarray}
\label{LIQP7}
    \dot{{A}} (t) &=& -ih^{S}{A}(t) - ih^{SB}{B}(t), \\
    \dot{B} (t) &=& -ih^{B}B(t) - ih^{SB \dagger}A(t).
    \label{eq:bathL}
\end{eqnarray}
We first solve the bath equations in Eq.~(\ref{eq:bathL}) and then substitute the solution to the system's equation of motion in Eq.~(\ref{LIQP7}). The formal solution of Eq.~(\ref{eq:bathL}) is given by,
\begin{equation}
\begin{aligned}
    b_{i} (t) &= i \sum_{r=1}^{L_{B}} \Big[g^{+}  (t - t_{0})\Big]_{i r} b_{r} (t_{0}) \\
    & \qquad  + \int_{t_{0}}^{\infty} d\tau \sum_{r=1}^{L_{B}} \sum_{s=1}^{L}  \Big[g^{+}  (t - \tau)\Big]_{i r} h^{SB \dagger}_{rs} a_{s}(\tau)
\end{aligned}
\end{equation}
where $i=1,2, \cdots L_{B}$ denotes the indices for bath operators. The Green's function 
\begin{equation}
\label{eq:g+t}
    g^{+} (t) = -i \,\theta(t)\, e^{-i h^{B} t}
\end{equation} 
is the solution of the homogeneous part of the Eq.~(\ref{eq:bathL}) and $\theta(t)$ is the Heaviside step function. Substituting this solution in Eq.~(\ref{LIQP7}), we obtain the quantum Langevin equation (QLE) for the system operators as,
\begin{equation}
\begin{aligned}
\label{TBPI1}
    \dot{a}_{i} (t) &= -i \sum_{s=1}^{L} h^{S}_{is} a_{s} (t) - i \, \eta_{i} (t) \\
    & \qquad - i \int_{t_{0}}^{\infty} d\tau \sum_{s=1}^{L}  \, \Sigma^{+}_{i s}(t-\tau) \, a_{s}(\tau)  
\end{aligned}
\end{equation}
Note that, in Eq.~(\ref{TBPI1}), the effect of the bath appears as a self-energy and a noise term, which are given respectively as
\begin{eqnarray}
&& \Sigma^{+}(t-\tau) = h^{SB} \, \Big[g^{+}  (t - \tau)\Big] \,h^{SB \dagger}, \\
&& \eta(t) = i \, h^{SB} \, \Big[g^{+}  (t - t_{0})\Big] \, B (t_{0}).
\end{eqnarray}
The statistical property of the noise operator gets determined by the initial condition of the bath density operator, as given in Eq.~(\ref{eq:rhoth}). As a result, $\langle \eta_i(t) \rangle =0$. The noise correlation at different times can be expressed in terms of the normal modes of the bath as
\begin{widetext}
\begin{equation}
\label{eq:etacorr}
    \Big \langle \eta^{\dagger}_{i} (t) \eta_{j} (t') \Big \rangle = \sum_{r,r'=1}^{L_{B}} h^{SB *}_{i r} \left( \sum_{q=1}^{L_{B}} U^{*}_{r q} U_{r' q} \bar{n}(\lambda^{B}_{q}) e^{i \lambda^{B}_{q} (t - t')} \right) h^{SB^{T}}_{r' j} \theta(t - t_{0}) \theta(t' - t_{0}) 
\end{equation}
\end{widetext}
where recall that, the matrix $U$ is responsible for  diagonalizing the single particle Hamiltonian $h^B$ of the bath [see Eq.~(\ref{LIQP4})]. Since we are interested in the steady-state limit, we first take $L_{B} \to \infty$ and then let $t_{0} \to - \infty$. As a result, $\theta(t - t_{0})$ and $\theta(t' - t_{0})$ in Eq.~(\ref{eq:etacorr}) are always equal to unity. Let us now define Fourier transformation of $\eta_i(t)$ as, 
\begin{equation}
\label{eq:etaFw}
    \tilde{\eta}_i(\omega)= \int_{-\infty}^{\infty} dt\,  e^{i \omega t}\, \eta_i(t)
\end{equation}
and the corresponding inverse is given as
\begin{equation}
    {\eta}_i(t)= \int_{-\infty}^{\infty} \frac{d\omega}{2\pi}\,  e^{-i \omega t}\, \tilde{\eta}_i(\omega). 
\end{equation}
Using Eq.~(\ref{eq:etaFw}) and Eq.~(\ref{eq:etacorr}), we get \begin{equation}
\label{LIQP9}
\begin{aligned}
\Big \langle \tilde{\eta}^{\dagger}_{i} (\omega) \tilde{\eta}_{j} (\omega')\Big \rangle &= 4 \pi^{2}\, \Gamma_{ji} (\omega)\, \bar{n}(\omega) \, \delta(\omega - \omega')\\
\end{aligned}
\end{equation}
where,
\begin{equation}
\begin{gathered}
    \Gamma_{ij} (\omega) = \sum_{r, r'=1}^{\infty} h^{SB}_{j r'} \rho_{r' r} (\omega) h^{SB \dagger}_{r i}, \\
    \rho_{r' r} (\omega)= \sum_{q=1}^{\infty} U^{*}_{r q} U_{r' q} \delta(\omega - \lambda^B_{q}).
\end{gathered}
\end{equation}
We obtain the solution of Eq.~(\ref{TBPI1}) in the Fourier space as
\begin{equation}
\label{LIQP8}
\begin{aligned}
    \tilde{a}_{i} (\omega) &= \sum_{s=1}^{L} G^{+}_{i s} (\omega)\, \tilde{\eta}_{s} (\omega)
\end{aligned}
\end{equation}
where $\tilde{a}_i (\omega)$ is the Fourier transformation of $a_i(t)$ with definition same as Eq.~(\ref{eq:etaFw}). The retarded Green's function $G^{+}(\omega)$ that appears in Eq.~(\ref{LIQP8}) is given as 
\begin{equation}
    G^{+} (\omega) = \Big[\omega I - h^{S} - \Sigma^{+}(\omega)\Big]^{-1}, 
\end{equation}
where $\Sigma^{+}(\omega)$ is now the self-energy matrix in the Fourier space and defined as,
\begin{equation}
 \Sigma^{+} (\omega) = h^{SB} \tilde{g}^{+} (\omega) h^{SB \dagger}.
 \end{equation}
Here, $\tilde{g}^{+} (\omega)$ is the Fourier transform of $g^{+} (t)$ [see Eq.~(\ref{eq:g+t})] and is given as
\begin{equation}
    \Big[\tilde{g}^{+} (\omega)\Big]_{r r'} = \sum_{q=1}^{\infty} \frac{U_{r q} U^{*}_{r' q}}{\omega - \lambda^B_{q} + i 0^{+}}
\end{equation}
where the small imaginary component appears to preserve the causality of $g^{+}(t)$.
Finally, using Eq.~(\ref{LIQP8}) and the noise-noise correlation in Eq.~(\ref{LIQP9}), we obtain the spatial local density in the steady state as:
\begin{eqnarray}
\label{TBPI3}
n_i(t) &=& \langle a^{\dagger}_{i}(t) a_{i}(t) \rangle \nonumber \\
&=& \int_{-\infty}^{\infty} d \omega \Big[G^{+}(\omega) \Gamma (\omega) G^{-}(\omega)\Big]_{ii}\, \bar{n}(\omega), 
\end{eqnarray}
where we recall that $\bar{n}(\omega)$ is either the Bose or the Fermi function as defined in Eq.~(\ref{eq:nw}). The integral in Eq.~(\ref{TBPI3}) can be performed numerically and the steady state occupation at each site can be determined exactly. Note that, the total occupation in the steady state can be obtained by following Eq.~(\ref{eq:totalN}). 

Having described the four methods in Sec.~(\ref{M1},\ref{M2}, \ref{M4}, and \ref{M3}) we now present our numerical findings using these methods. 
\begin{figure}
    \centering
    \includegraphics[width=1.0\columnwidth]{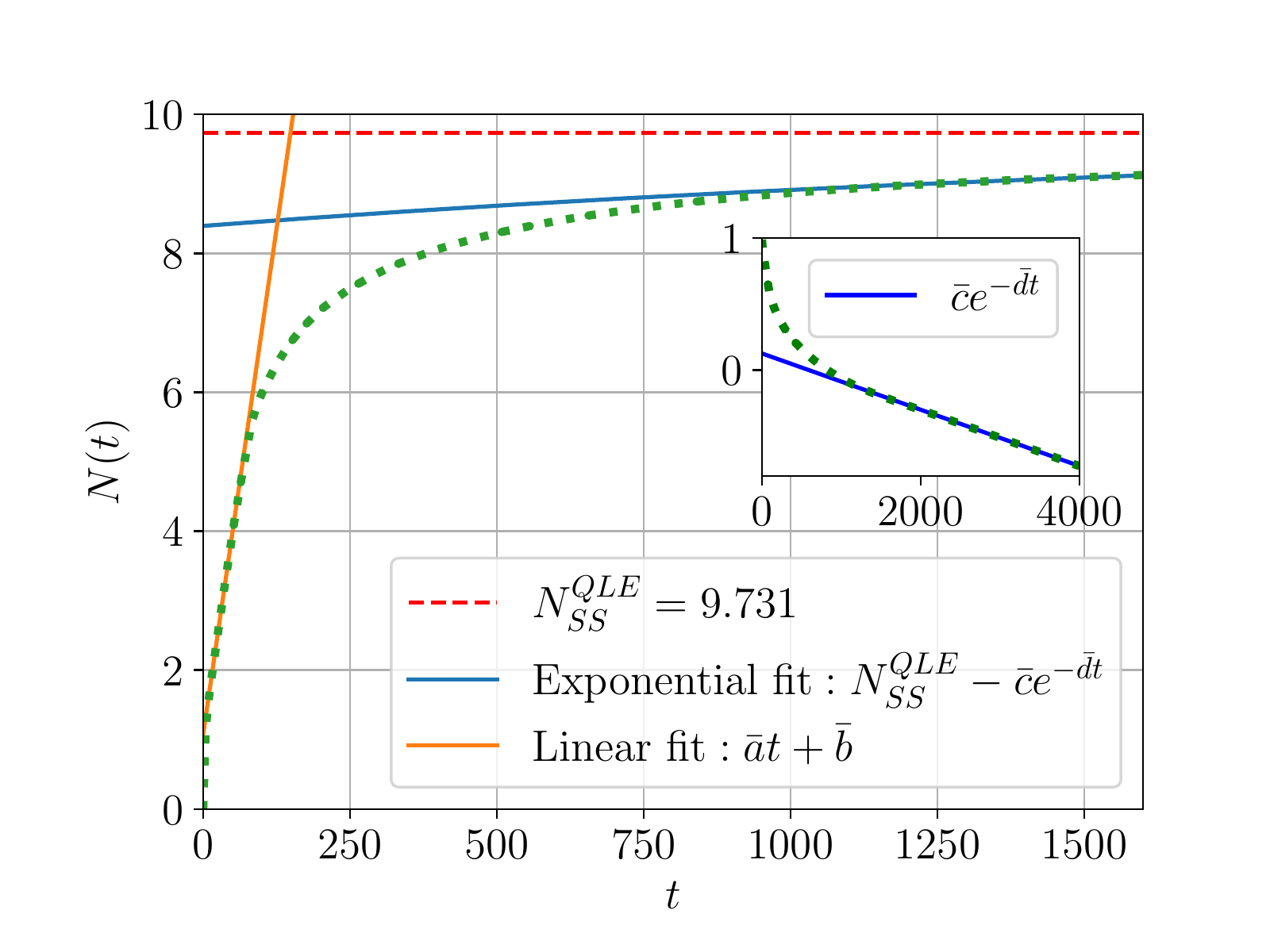}
    \caption{Behaviour of total occupation $N(t)$ [Eq.~\eqref{eq:totalN}]  versus $t$ for system size $L=40$ for the bosonic case, using direct exact numerics as described in Sec.~(\ref{M1}). The bath is connected at site number $m=21$. The linear early time growth and exponential late time saturation can be clearly seen. Red dashed line shows the steady state value $N^{QLE}_{SS}$ obtained from exact quantum Langevin equation approach, described in Sec.~(\ref{M3}).  The parameters are $g = 0.5$, $ \gamma = 1$, $t_{B} = 1$, $\beta = 1$, and $\mu = -2.01$. Note that the fitting parameters for the early time linear growth $\bar{a} t + \bar{b}$ are $\bar{a} = 0.058$, $\bar{b} = 1.082$. For the late time behaviour, the fitting parameters for the exponential relaxation $N^{QLE}_{SS}- \bar{c} \, e^{-d t}$ are  $\bar{c} = 1.359$, $\bar{d} = 4.95 \times 10^{-4}$ with $N^{QLE}_{SS}$ chosen to be same as the steady state value obtained from QLE ($N_{SS}^{QLE}=9.731)$. This implies that for $L=40$, the time scale to reach steady state is $t_{SS} = 1/\bar{d} \sim 2000$.
    The inset shows  the plot for $\big[\log\big(N^{QLE}_{SS} \!-\! N(t)\big)\big]$ vs $t$ (green dots) and it clearly demonstrates the long time exponential relaxation in time towards the steady state. We choose $\gamma=1$ (non-perturbative regime in system-bath coupling) to ensure that the system relaxes towards steady-state relatively fast. For this figure, the system size $L=40$ is chosen such that both the early time linear behaviour and long time exponential behaviour are clearly visible.}
    \label{f2}
\end{figure}

\begin{figure}
    \centering
    \includegraphics[width=1\columnwidth]{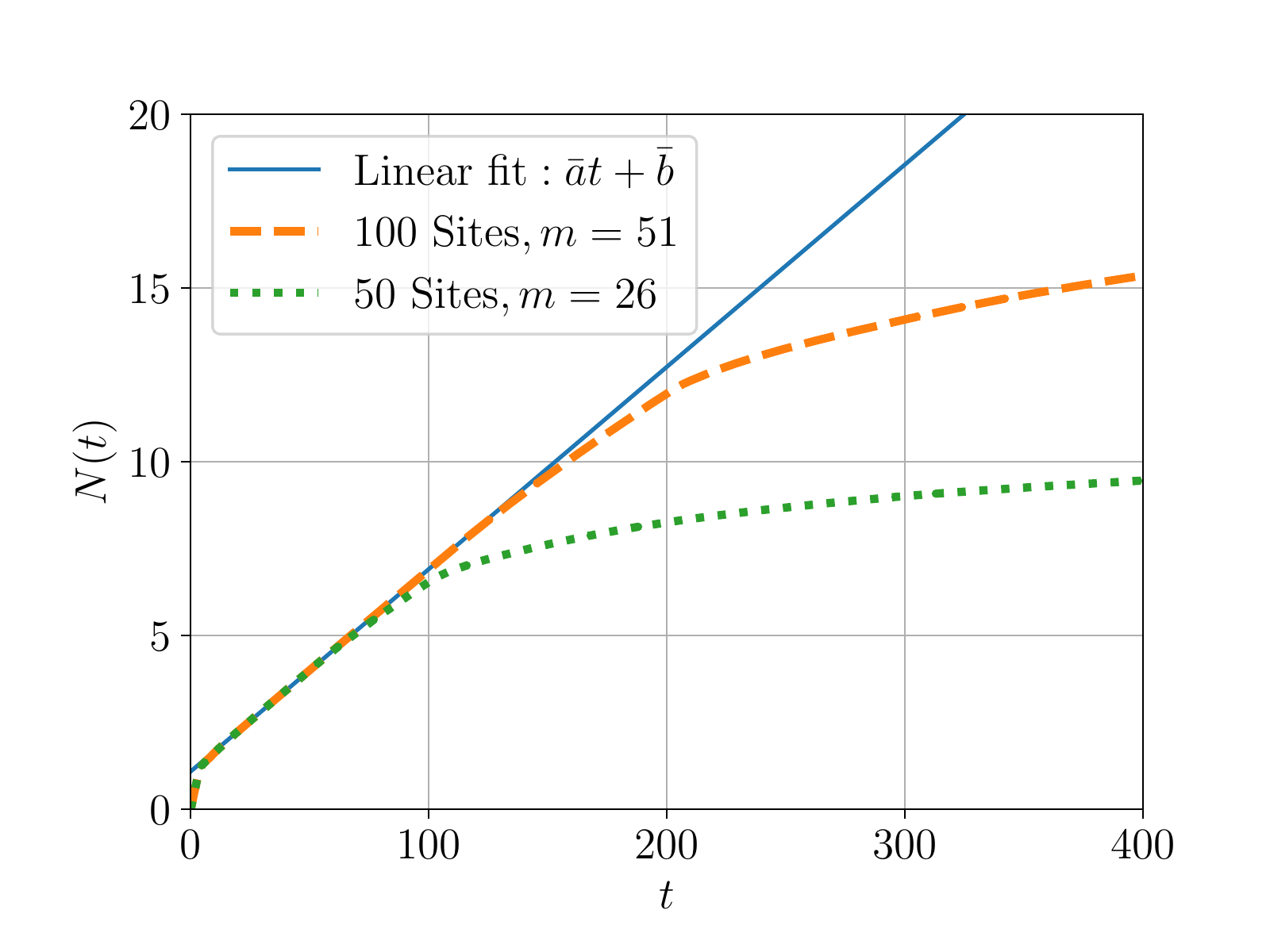}
    \caption{Total occupation $N(t)$ [Eq.~\eqref{eq:totalN}] versus $t$ for different system sizes using direct numerics as described in Sec.~(\ref{M1}). The parameters used are $g = 0.5$, $t_{B} = 1$, $\beta = 1$, $\mu = -2.01$, and $ \gamma = 1$ are exactly the same as in Fig.~(\ref{f2}) except the system size. It can be seen that the deviation from the linear growth starts at a time scale that scales with the system size $L$. The fitting parameters are $\bar{a} = 0.058$, $\bar{b} = 1.082$.}
    \label{f3}
\end{figure}

\begin{figure}
    \centering
    \includegraphics[width=1\columnwidth]{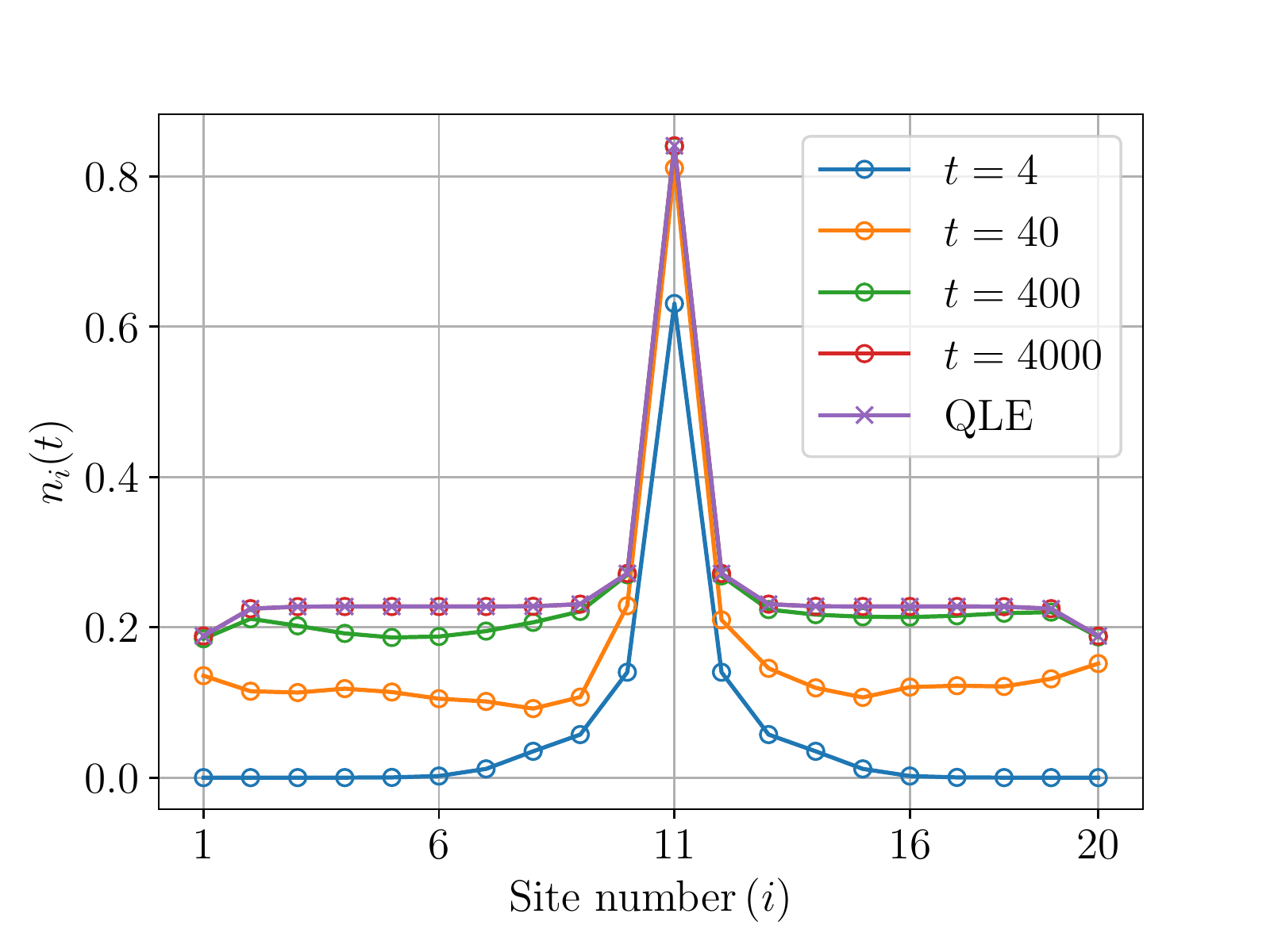}
    \caption{Local density profile $n_i(t)$ [Eq.~(\ref{eq:spatial})] for $L=20$ bosonic lattice sites with bath attached at a particular site ($m = 11$) for various time snapshots, using direct numerics (empty circles), as discussed in Sec.~(\ref{M1}). The long time limit of this density profile agrees perfectly with that obtained from QLE (cross), as discussed in Sec.(\ref{M3}). The parameters are $t_{B} = 1$, $g = 0.5$, $\beta = 1$, $\mu = -2.01$ and $ \gamma = 1$. Note that apart from the system size the parameters chosen here are exactly the same as in Fig.~(\ref{f2}) and Fig.~(\ref{f3}). The system size $L=20$ is chosen keeping in mind computational feasibility and to ensure a relatively quick approach to steady-state.}
    \label{f4}
\end{figure}

\begin{figure}
    \centering
        \includegraphics[width=\columnwidth]{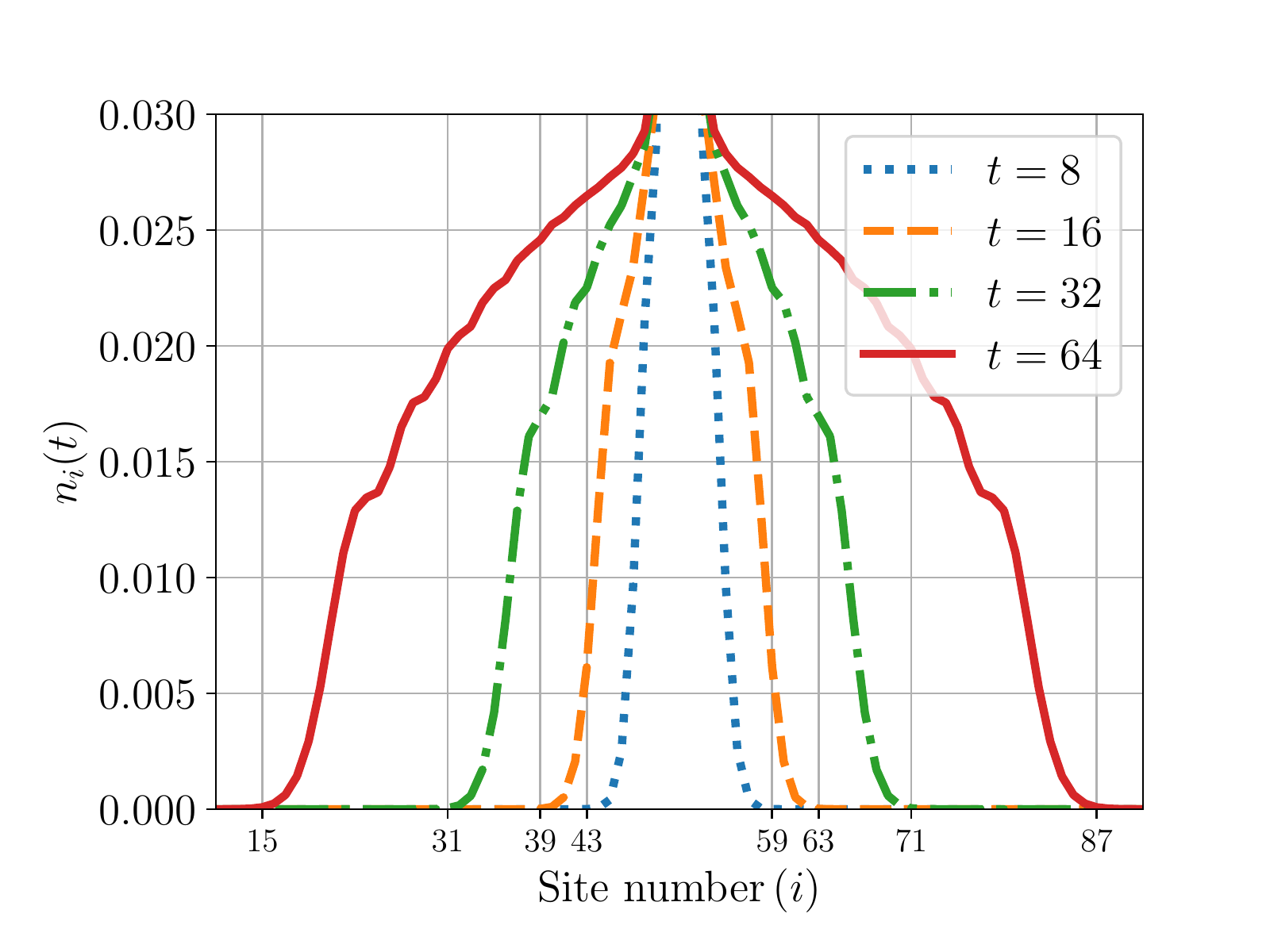}
        \caption{Local density profile $n_i(t)$ [Eq.~(\ref{eq:spatial})] for $L=100$ bosonic sites with bath attached at $m = 51$ from direct exact numerics, described in Sec.~(\ref{M1}). We choose relatively short times to clearly demonstrate ballistic growth of the density profile. We truncate the $y-$ axis to highlight the propagation of the density front. Note that this front of the density profiles spreads with velocity $c=2\,g$, where $g$ is the inter-site hopping within the lattice. This figure clearly indicates the presence of scaling which is demonstrated in Fig.~(\ref{f5.2}). The parameters are $t_{B} = 1$, $g = 0.25$, $\beta = 1$, $\mu = -2.01$ and $ \gamma = 1$. Note that the parameters chosen here are exactly the same as in Fig.~(\ref{f2}), Fig.~(\ref{f3}), and Fig.~(\ref{f4}) except the value of $g$. $g=0.25$ is chosen here to in order to illustrate ballistic spreading over a computationally feasible system size $L$.}
        \label{f5.1}
\end{figure}

\begin{figure}
    \centering
    \includegraphics[width=1\columnwidth]{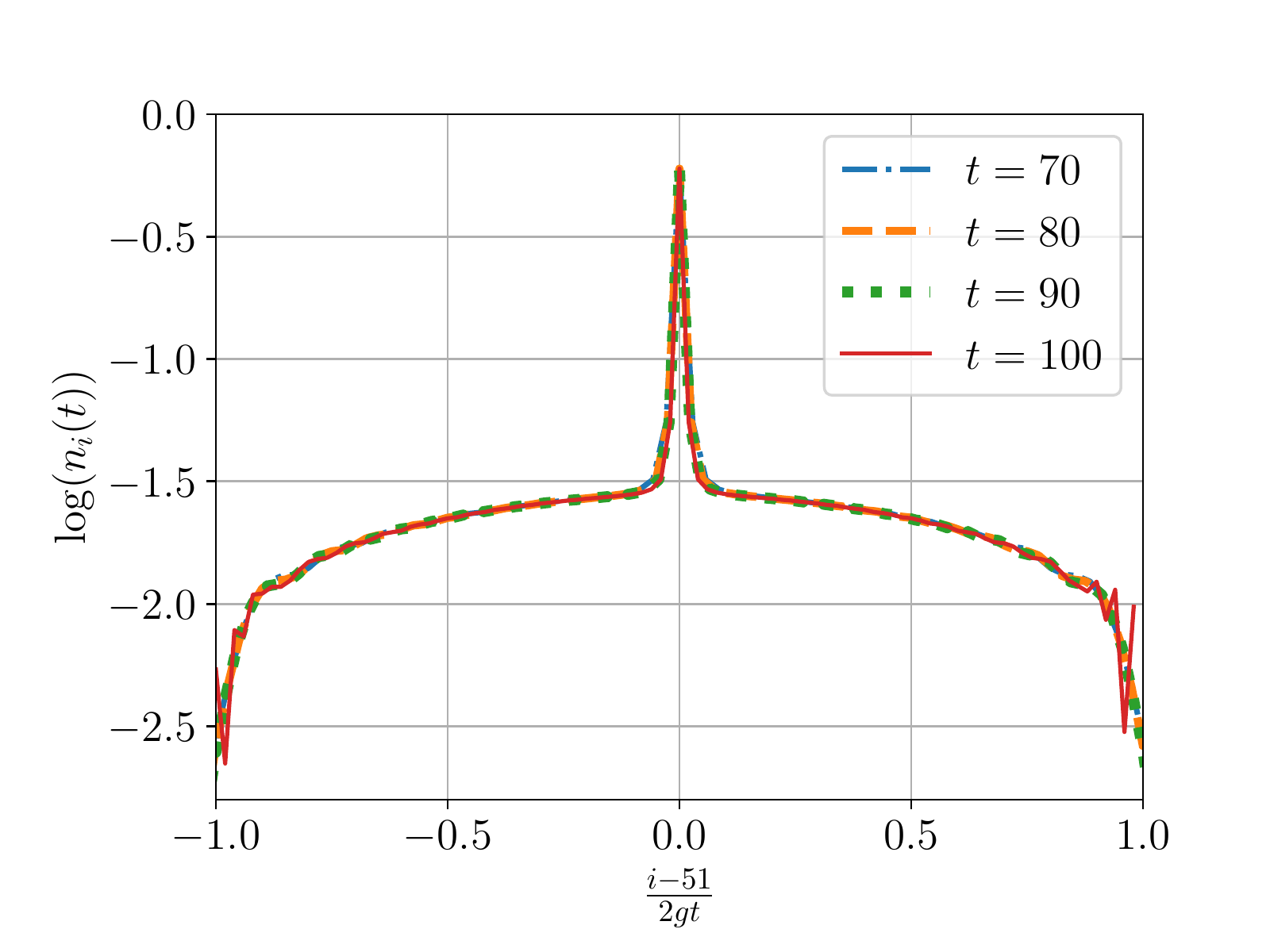}
    \caption{A scaled plot of $\log\big(n_{i} (t)\big)$ [Eq.~(\ref{eq:spatial})] for $L=100$ bosonic sites with bath attached at $m = 51$ to demonstrate the ballistic spread of spatial density profile $n_i(t)$ [Eq.~(\ref{eq:spatial})]. 
    The parameters are $t_{B} = 1$, $g = 0.25$, $\beta = 1$, $\mu = -2.01$, and $ \gamma = 1$.}
    \label{f5.2}
\end{figure}

\begin{figure}
    \centering
    \includegraphics[width=1\columnwidth]{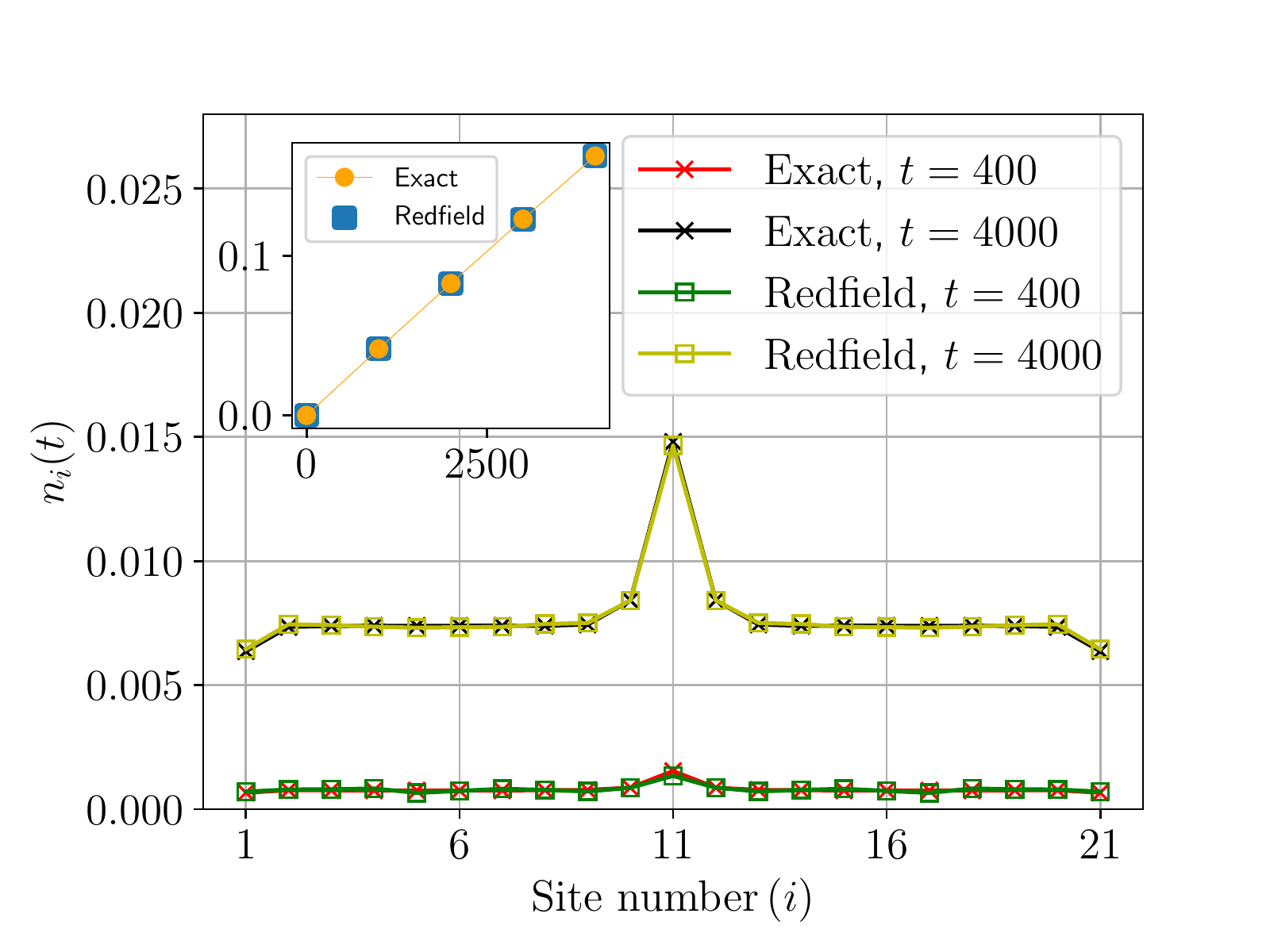}
    \caption{Local density profile $n_i(t)$ [Eq.~(\ref{eq:spatial})] for $L=21$ bosonic lattice sites with bath attached at a particular site ($m = 11$) for various time snapshots ($t=400$, $t=4000$) using direct exact numerics (cross), as discussed in Sec.~(\ref{M1}). We show good agreement of these results with that obtained using Redfield equation approach (squares), discussed in Sec.~(\ref{M2}). 
    The inset shows total occupation $N(t)$ [Eq.~\eqref{eq:totalN}] for $t=1000,2000,3000,4000$ with both direct exact numerics [Sec.~\ref{M1}] and Redfield [Sec.~\ref{M2}] approach. The slope obtained from this plot perfectly matches with the slope extracted following Eq.~\eqref{eq:early-time}. The parameters are $t_{B} = 1$, $g = 0.5$, $\beta = 1$, $\mu = -2.01$ and $ \gamma =0.01$. Note that as $\gamma=0.01$ and $g=0.5$, the Redfield approach is well suited for comparison with the direct exact numerics. Also as a consequence of  low value of $\gamma$, the time to reach steady state is very long and much higher than the times presented in this figure. This further implies that the steady value of local density is far from the values presented here.}
   \label{f4a}
\end{figure}

\section{Numerical Results}
\label{sec:numerical_results}

In this section we present our numerical results for a one-dimensional nearest neighbour tight binding lattice which is coupled to an equilibrium bath at a particular site [recall Fig.~(\ref{schematic})]. The quantities of interest are (i) local occupation number $n_{i} (t)$  versus $i$ at  fixed time snapshots and (ii) the total occupation $N(t)$ versus $t$. We will mainly focus on presenting results for the bosonic case. We will briefly discuss the fermionic case and  highlight interesting similarities and differences between the two.

Unless otherwise stated, we choose the following parameters  for the simulations. For the bath, we fix the parameters as, inter-site hopping $t_{B} = 1$, chemical potential $\mu = -2.01$, and inverse temperature $\beta = 1$. For the direct numerics, discussed in Sec.~(\ref{M1}), we always choose $L_{B}=4096$. 
We connect the bath at a particular site $m$ of the lattice. 
For the Redfield [Sec.~(\ref{M2})], Lindblad [Sec.~(\ref{M4})], and quantum Langevin equation approach [Sec.~(\ref{M3})], the bath is considered to be a semi-infinite one-dimensional tight binding chain, and the corresponding form of the spectral density $J(\omega)$ can be obtained exactly, given as \cite{purkayastha2016out}
\begin{equation}
\label{eq:jw_sqrt}
J(\omega) = \frac{2 \gamma^{2}}{t_{B}} \sqrt{1 - \frac{\omega^{2}}{4\, t^{2}_{B}}}.  
\end{equation}
Before proceeding further, we note that depending on the method employed and specific quantities of interests, the system size $L$, hopping parameter $g$, and the system-bath coupling $\gamma$ are chosen by taking into account computational feasibility and better clarity of presentation. 

\subsection{Non-perturbative regime in system-bath coupling}
\label{sec:NPR}
In Fig.~(\ref{f2}) we show the total occupation $N(t)$ as a function of time $t$ using the direct exact numerics, described in Sec.~(\ref{M1}). The early time linear behaviour and long time exponential relaxation towards the steady state is clearly seen. We also present the steady state value obtained from the quantum Langevin equation approach, described in Sec.~(\ref{M3}) and observe that the long-time limit for $N(t)$ from direct exact numerics approaches to the exact steady state value. 
Notice that we clearly observe an exponential relaxation of $N(t)$ towards the steady state [see inset of Fig.~(\ref{f2})]. This is consistent with relaxation dynamics of finite size systems coupled to a generic bath. This exponential relaxation can, in fact, be rigorously established following the Redfield approach [Sec.~\ref{M2}].  $N(t)$ reaching a steady state value is a result of finite system size which here is taken to be $L=40$. Also, to ensure that steady-state value is reached relatively fast, we choose $\gamma=1$ which falls into the non-perturbative regime of the system-bath coupling. We expect that the time to reach steady state $t_{SS}$ increases with system size $L$. By performing direct exact numerics we find that the time to reach steady state  for $L=16$, $L=20$, and $L=40$ are $t_{SS} \sim 175, 330$, and $2000$, respectively. Based on these numbers, we find that the dependence of $t_{SS}$ on system size $L$ is $t_{SS} \sim L^{\delta}$ where $\delta \approx 2.5$. In the large $L$ limit one would expect the scaling to go as $t_{SS}\sim L^2$ which is related to the fact that the adjacent energy gaps in a tight-binding chain scale as $1/L^2$.

One would expect $N(t)$ to grow linearly for an infinite lattice. To demonstrate this, using direct numerics [Sec.~(\ref{M1})], we show in Fig.~(\ref{f3}) the behaviour of $N(t)$ versus $t$ for different system sizes. It can be noticed that the deviation from the linear growth starts at a time scale that scales with the system size $L$.

In Fig.~(\ref{f4}), we show the spatial density profile $n_i(t)$ [Eq.~(\ref{eq:spatial})] as a function of lattice coordinate $i$ for various time snapshots $t$ using direct numerics [Sec.~(\ref{M1})]. The long-time limit of this density profile agrees perfectly with that obtained using the quantum Langevin equation approach, describe in Sec.~(\ref{M3}). Note that here we choose the lattice size $L=20$ and the bath is connected to the lattice at site $m=11$. Therefore this particular site shows maximum average local occupation and eventually thermalizes with the bath, thereby settling to a finite value. The nearby sites gradually develop local occupation and finally settle down to a finite value, owing to indirect thermalization with the bath. In the long-time limit, an interesting pattern of local density profile is formed. At very large times, the relatively flat pattern of the local density profile, away from the center, is an interesting observation. Note that apart from the system size the parameters chosen in Fig.~(\ref{f4}) are exactly the same as in Fig.~(\ref{f2}) and Fig.~(\ref{f3}) and therefore we remain in the non-perturbative system-bath coupling regime. We choose system size $L=20$ keeping in mind computational feasibility and to ensure a relatively quick approach to steady-state.

In Fig.~(\ref{f5.1}) we present a zoomed view of the spread of local density profiles $n_i(t)$ for $L=100$ sites with bath connected at $m=51$, for different time snapshots. The ballistic spread of the density profile with velocity $2\,g$ can be clearly seen ---where we recall that $g$ is the inter-site hopping within the lattice system. This indicates a scaling form for the profile which is presented in Fig.~(\ref{f5.2}). 
Note that the parameters chosen in Fig.~(\ref{f5.1}) and Fig.~(\ref{f5.2}) are exactly the same as in Fig.~(\ref{f2}), Fig.~(\ref{f3}), and Fig.~(\ref{f4}) except the value of $g$ which is chosen to be $0.25$ in order to illustrate ballistic spreading over a computationally feasible system size $L$.
\\

\subsection{Perturbative regime in system-bath coupling}
\label{sec:PR}
Next we discuss the regime of weak system bath coupling which further allows us to employ the Redfield [\ref{M2}] and Lindblad [\ref{M4}] approaches. Unlike Fig.~(\ref{f2}), Fig.~(\ref{f3}), and Fig.~(\ref{f4}) where we had set $\gamma=1$, in Fig.~(\ref{f4a}) we choose $\gamma=0.01$ to ensure that we remain in weak system-bath (perturbative) coupling regime. We retain the value of $g=0.5$ as before which therefore does not fall in the validity of local Lindblad equation approach, as was also mentioned in Sec.~(\ref{M4}). In Fig.~(\ref{f4a}) we first compare the local density profile $n_i(t)$ obtained following direct exact numerics, discussed in Sec.~(\ref{M1}) and the Redfield approach, discussed in Sec.~(\ref{M2}). We observe perfect agreement at various time snapshots. Moreover, the inset in Fig.~(\ref{f4a}) also shows excellent agreement between the two approaches for the total occupation $N(t)$ [Eq.~\eqref{eq:totalN}]. The slope obtained from this inset plot perfectly matches with the slope extracted following the short time (relative to the time to reach the steady state) dynamics described by Eq.~\eqref{eq:early-time}. Note that as a consequence of lower value of system-bath coupling $\gamma$, the time require to reach steady state is very long and much higher than the time snapshots presented in Fig.~(\ref{f4a}). This further implies that the steady value of local density is far from the values presented in Fig.~(\ref{f4a}).

\begin{figure}
    \centering
    \includegraphics[width=1\columnwidth]{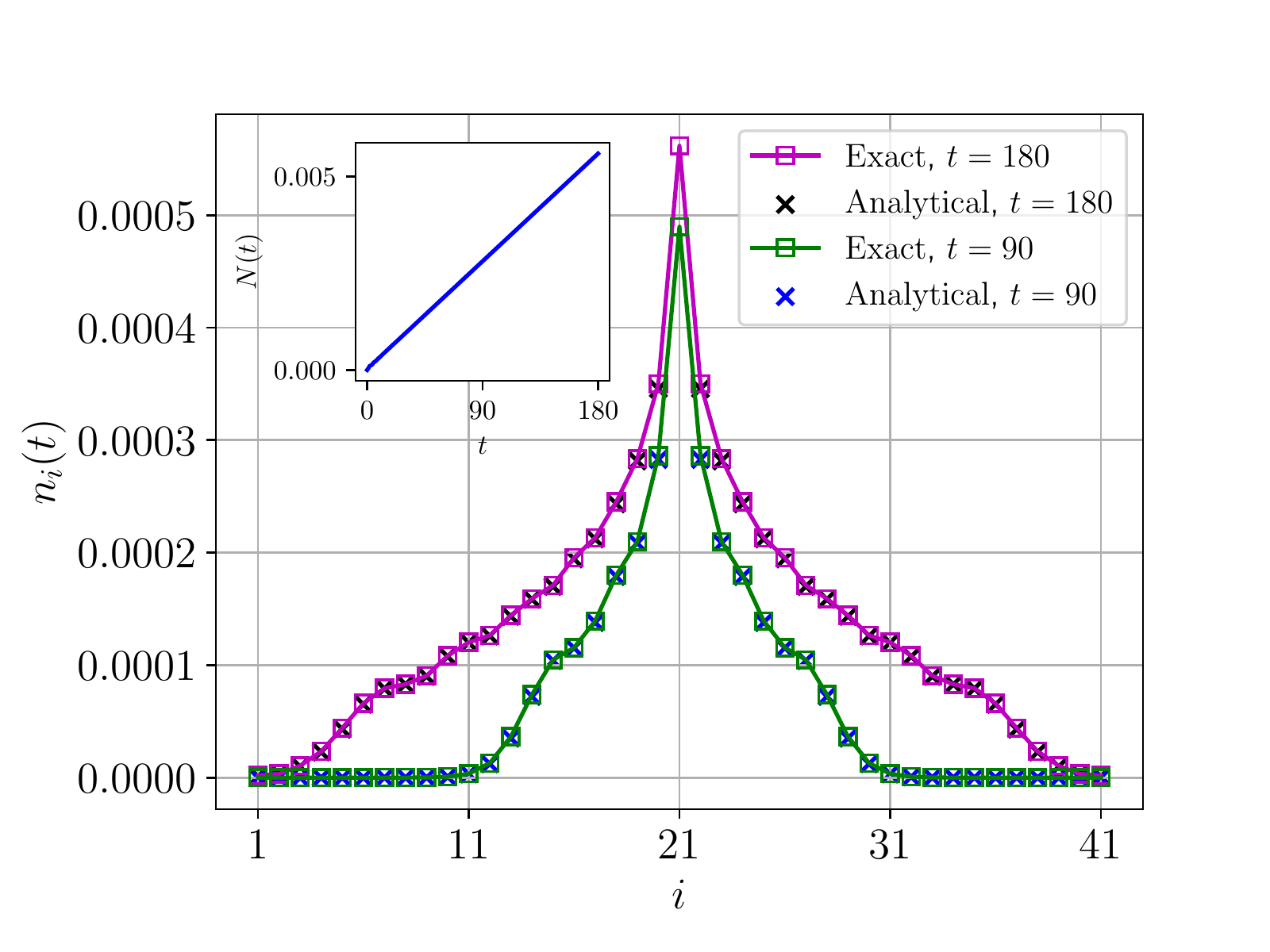}
    \caption{Spatial density profile  $n_i(t)$ [Eq.~(\ref{eq:spatial})] for bosons for $L=41$ sites with the bath attached at $m=21$ for two different time snapshots following direct exact numerics [Sec.~(\ref{M1})], and analytical result based on Lindblad approach [Eqs. \eqref{eq:ni_int_exact-m} and  \eqref{eq:scaledS-m}] discussed in Sec.~(\ref{M4}). We notice excellent agreement between the two approaches. The parameters chosen are $t_{B} = 1$, $g = 0.05$, $\beta = 1$, $\mu = -2.01$ and $ \gamma =0.01$. The plots are shown for time snapshot $t=90$ and $t=180$. Note that for the analytical case, no data is presented at $m=21$ as analytical expression in Eq.~\eqref{eq:scaledS-m} is expected to only hold away from the lattice site where the bath is attached. Recall that we have chosen weak inter-site hopping $g$ to ensure the validity of local Lindblad approach. The inset shows the plot for total occupation $N(t)$ vs $t$ for the same parameter values. The slope in the inset preciously turns out to be $2 \, \Gamma_G$ and therefore matches with the prediction  in Eq.~\eqref{eq:N_g_large} [Sec.(\ref{M4})].}
    \label{f8}
\end{figure}

\begin{figure}
    \centering
    \includegraphics[width=1\columnwidth]{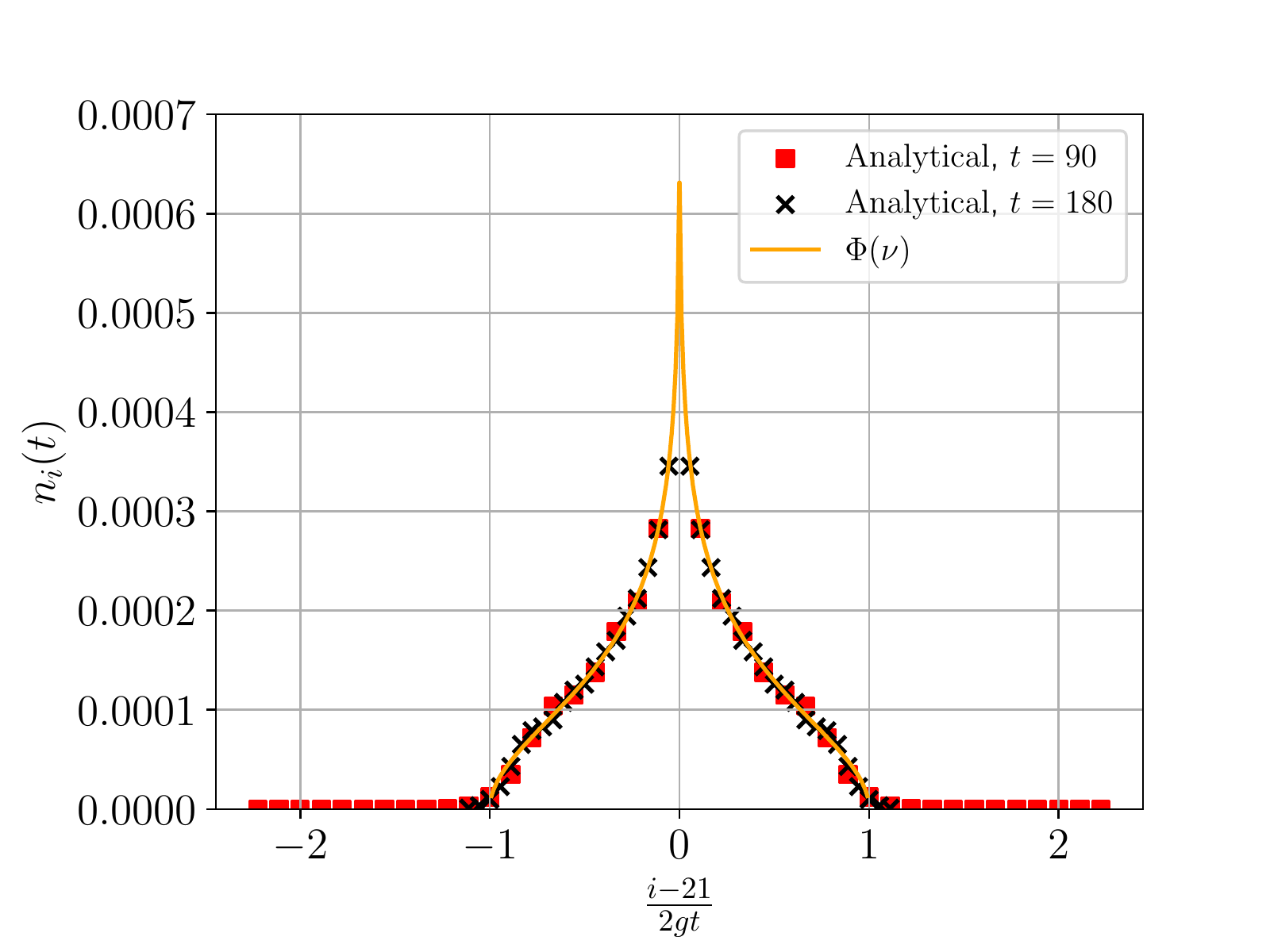}
    \caption{Spatial scaled density profile  $n_i(t)$ [Eq.~(\ref{eq:spatial})] for bosons for $L=41$ sites with the bath attached at $m=21$ for two different time snapshots following analytical result based on Lindblad approach Eqs.~\eqref{eq:ni_int_exact-m} and \eqref{eq:scaledS-m} and the scaling form Eq.~\eqref{eq:ni_scale-main-2} discussed in Sec.~(\ref{M4}). We notice excellent agreement between the two results. The parameters chosen are exactly same as those of Fig.~(\ref{f8}) i.e.,  $t_{B} = 1$, $g = 0.05$, $\beta = 1$, $\mu = -2.01$ and $ \gamma =0.01$. The plots are shown for time snapshot $t=90$ and $t=180$. Recall that we have chosen weak inter-site hopping $g$ to ensure the validity of Lindblad approach.}
    \label{f10}
\end{figure}

Next, we further reduce the value of inter-site hopping parameter $g$ to $g=0.01$. This enables us to be in a regime where the local Lindblad approach is valid. In Fig.~(\ref{f8}) we plot the local density $n_i(t)$ for two different time snapshots and demonstrate excellent agreement between the analytical results given by Eqs. \eqref{eq:ni_int_exact-m} and  \eqref{eq:scaledS-m} with direct exact numerics [Sec.~(\ref{M1})]. The inset in Fig.~(\ref{f8}) shows a plot for $N(t)$ vs $t$ in the same parameter regime which shows a perfect linear growth with slope $2 \Gamma_G$ and therefore matches with the prediction in Eq.~\eqref{eq:N_g_large} 
In Fig.~(\ref{f10}) we use the same parameters as in Fig.~(\ref{f8}) and demonstrate excellent agreement between the scaled version of data in Fig.~(\ref{f8}), analytical scaling form [Sec.~(\ref{M4})] given in Eq.~\eqref{eq:ni_scale-main-2}.

\begin{figure}
\centering
\includegraphics[width=\columnwidth]{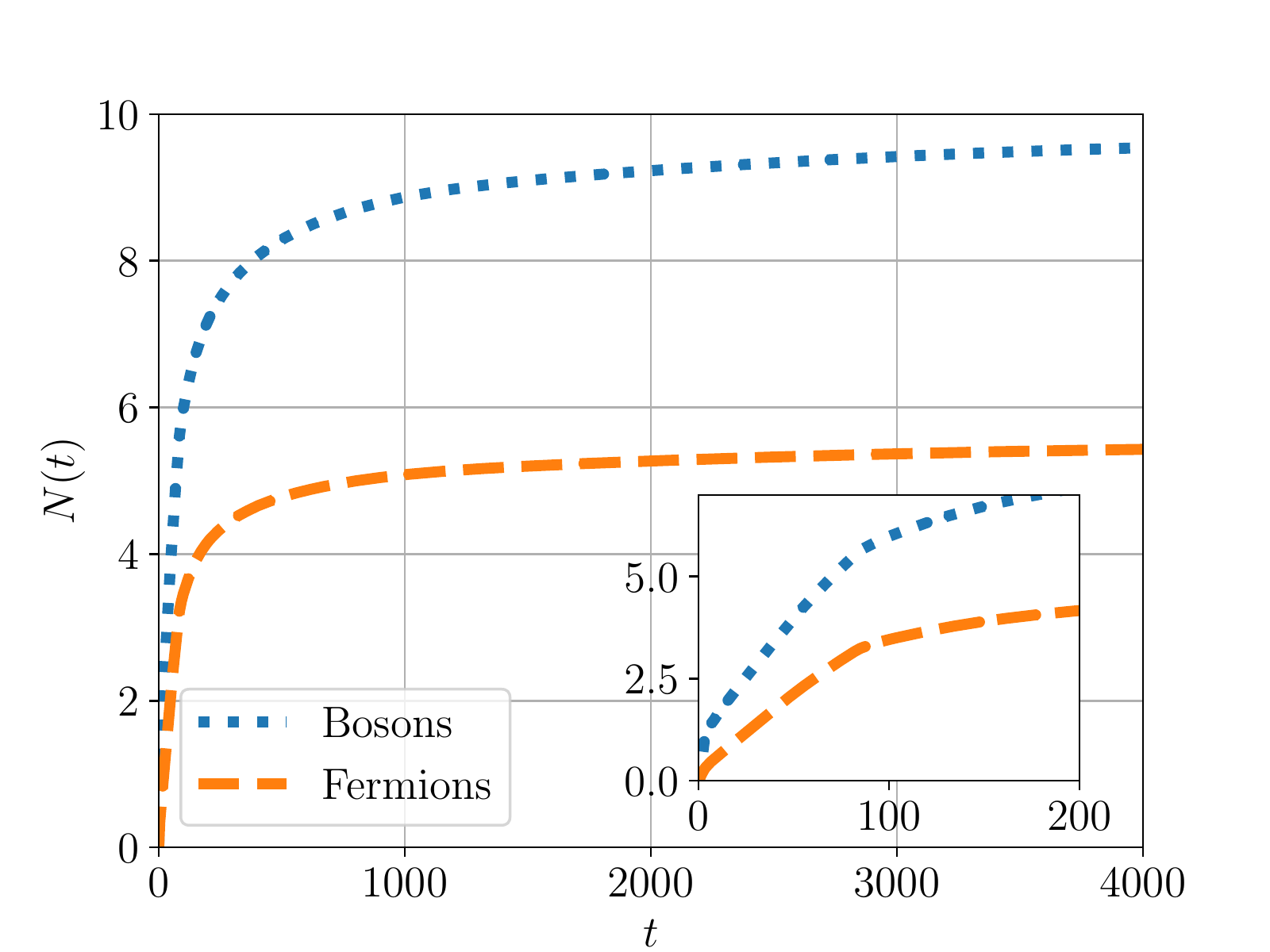}
\caption{Comparison between fermions and bosons for total occupation $N(t)$ for $L=40$ sites with bath attached at $m = 21$, following direct exact numerics discussed in Sec.~(\ref{M1}). It is clear from the plot that the eventual saturation value for fermions is smaller than that of bosons. The inset denotes the zoomed version of the dynamics at relatively short times and clearly shows that rate of growth for fermionic case is smaller than that of bosonic case. The parameters chosen are $t_{B} = 1$, $g = 0.5$, $\beta = 1$, $\mu = -2.01$ and $ \gamma = 1$. Note that although we have presented the results only from direct exact numerics (Sec.~\ref{M1}), similar trends can be obtained following other methods.}
\label{compare-fb}
\end{figure}

So far we presented results for bosons. We now briefly make a few comments about fermions highlighting the similarities and differences. We find that owing to Pauli exclusion principle, fermions experience blockade which makes its quantum dynamics  different from that of bosons. 
The rate of growth of $N(t)$ at small times is higher for bosons as they are not limited by Pauli exclusion principle obeyed by fermions. As a result of such slow growth for fermions, in the long time limit overall total occupation $N(t)$ within the lattice is significantly lower in comparison with bosons. Similar to bosons, the fermions also exhibit overall early time linear growth and exponential relaxation at long-times. In Fig.~(\ref{compare-fb}) we demonstrate these trends following the direct exact numerics described in Sec.~(\ref{M1}). In suitable parameter regimes one can employ the other methods [Sec.~(\ref{M2}), Sec.~(\ref{M4}), Sec.~(\ref{M3})] and notice similar trends.

\section{Discussions and Comparisons with Previous Works}
\label{sec:comparison}
As mentioned earlier, local gain or loss experienced by a system owing to a connection to a reservoir is an actively investigated area of research. Therefore we place our work in the context of certain recent works. Note that, in our setup, the bath is responsible for simultaneous injection and/or removal of bosons or fermions with rates that obey a detailed balance condition. In other words, the rate of injection and removal are not arbitrary but are related. This condition is respected by all the methods [Sec.~(\ref{sec:setup})] discussed in our work. If one considers only pure injection process (thereby not obeying detailed balance condition) both short and long-time quantum dynamics can be drastically different \cite{krapivsky2019free,krapivsky2020free}. For the case of bosons interesting dynamical transitions have been reported \cite{krapivsky2020free} when the system is subjected to incoherent pump with no loss channels. More preciously, in the limit of large system sizes, $N(t)$ in Eq.~(\ref{eq:totalN}) exhibits exponential/power law growth depending on the incoherent pumping strength. However, an analogous setup for fermions \cite{krapivsky2019free} does not display such dynamical transitions and $N(t)$ grows linearly with $t$. This is an example where Pauli exclusion or lack thereof can have remarkably different consequences. 

It is easy to note that, if only incoherent pumping at a local site needs to be incorporated then this amounts to artificially setting the loss coefficient $\Gamma_L=0$ (thereby allowing the breakdown of detailed balance) in the Lindblad equation in Eq.~\eqref{eq:mod-Lind}.
This results in the following equation of motion for the two-point correlation functions $C_{i,j}$ as defined in Eq.~\eqref{eq:correlation} 
\begin{eqnarray}
\label{eq:corr-Lind}
    \frac{d C_{i,j}}{dt} &= i \, g\, (C_{i-1, j} \!-\! C_{i, j+1} \!+\! C_{i+1, j} \!-\! C_{i, j-1})\nonumber \\ 
    & \pm \Gamma_G (\delta_{i m} + \delta_{j m})\, C_{i, j} + 2\, \Gamma_G \delta_{m i} \delta_{m j}
\end{eqnarray}
where note the crucial sign difference in the second term with $+$ sign indicates bosons and $-$ sign indicates fermions.  This $\pm$ sign in Eq.~(\ref{eq:corr-Lind}) arising due to quantum statistics leads to crucial change in the dynamics for local density $n_i(t)$ in Eq.~\eqref{eq:spatial} and total occupation $N(t)$ in Eq.~\eqref{eq:totalN}. More explicitly, following the differential equation in Eq.~(\ref{eq:corr-Lind}), one can show a linear growth for $N(t)$ versus $t$ for fermions \cite{krapivsky2020free}, whearas for bosons  exponential/power law growth in time can be seen \cite{krapivsky2020free}. 

It is important to note that if one has to respect the detailed balance condition, it is not possible to set $\Gamma_L$ to 0. In fact, one can interestingly notice that from Eq.~(\ref{eq:GammaG}) and Eq.~(\ref{eq:GammaL}) that for bosons, $\Gamma_L > \Gamma_G$ is always true. As a consequence, the term $-(\Gamma_L \mp \Gamma_G) (\delta_{i m} + \delta_{j m})$ in Eq.~(\ref{eq:mod-corr-Lind}) is always negative for bosons. The same is straightforward to notice even for the case of fermions.  Therefore, for both bosons and fermions, Eq.~(\ref{eq:mod-corr-Lind}) bares strong structural resemblance with the equation obtained when a fermionic lattice [corresponding to negative sign in the second term of Eq.~\eqref{eq:corr-Lind}] is subjected only to incoherent pump \cite{krapivsky2019free} yielding  Eq.~(\ref{eq:corr-Lind}). This argument explains why via the methods employed in this work we see linear growth in $N(t)$ irrespective of whether one considers bosons or fermions.

\section{Summary and Outlook}
\label{sec:Conclusions}
In this work, we demonstrate 
how a complex interplay between unitary, non-unitary dynamics and quantum statistics can lead to non-trivial quantum dynamics and subsequent steady-state. We considered the 
setup when an empty lattice is locally connected to a reservoir [Fig.~(\ref{schematic})]. The main quantities of interest were local spatial density profile $n_i(t)$ [Eq.~(\ref{eq:spatial})] and the total occupation $N(t)$  [Eq.~(\ref{eq:totalN})]. We employed four methods --
(i) direct exact numerics for correlation matrix [Sec.~(\ref{M1})], (ii) Redfield equation [Sec.~(\ref{M2})], (iii) Lindblad equation [Sec.~(\ref{M4})], and (iv) exact quantum Langevin approach [Sec.~(\ref{M3})].

We showed that the initial growth for the total occupation $N(t)$ for both bosons and fermions is linear in time and it subsequently saturates (for a finite lattice size) to a constant value in an exponential manner. For infinite lattice there is no saturation and $N(t)$ grows linearly with time. The local spatial density profile $n_i(t)$ exhibits a ballistic spatial spread for both bosons and fermions. At any fixed lattice coordinate, $n_i(t)$ initially grows in time and eventually saturates owing to equilibration with the bath. Our simulation results indicate that the equilibration time $t_{SS} \sim L^{\delta}$  with $\delta \approx 2.5$. However, we expect $\delta=2$  in the large $L$ limit. Our work unravels the universal features and the differences between bosons and fermions the cause of which is rooted in quantum statistics. In the context of recent works on this subject, it is to be noted that our microscopic starting point is drastically different from phenomenological approaches \cite{krapivsky2019free,krapivsky2020free}. We show that, restoring detailed balance condition plays a pivotal role in deciding the fate of the quantum dynamics. We also show that our findings for spatial density profile obey 
analytical forms in an appropriate parameter regime.

Future work will be directed towards understanding quantum dynamics for filling particles in higher-dimensional lattices with arbitrary geometries and fully/partially connected networks \cite{wright2019benchmarking,xu2020probing,song2019generation,PhysRevApplied.16.024018,ray2022localization}. Understanding the full counting statistics \cite{PhysRevLett.110.060602} and distribution of total occupation, $P[N(t)]$, will be a problem of significant interest. 
A challenging and interesting question is understanding quantum dynamics and thermalization when reservoirs are locally connected to empty lattices that can host interacting bosons or fermions. It is important to highlight that with the current state-of-art experimental progress in absorption imaging techniques \cite{inguscio2008ultra,dalfovo1999theory,joseph2011observation} and quantum gas microscopy \cite{qm1,qm2,qm3}, it has now become feasible to measure local density profiles for systems with a very high precision even to the resolution at the level of a single atom.  

\begin{acknowledgments}
We thank K. Mallick for useful discussions. A.T. would like to thank the Long Term Visiting Students Program (LTVSP) of ICTS Bangalore. B. K. A. acknowledges the MATRICS grant (MTR/2020/000472) from SERB, Government of India and the Shastri Indo-Canadian Institute for providing financial support for this research work in the form of a Shastri Institutional Collaborative Research Grant (SICRG). M.K. would like to acknowledge support from the project 6004-1 of the Indo-French Centre for the Promotion of Advanced Research (IFCPAR), Ramanujan Fellowship (SB/S2/RJN-114/2016), SERB Early Career Research Award (ECR/2018/002085) and SERB Matrics Grant (MTR/2019/001101) from the Science and Engineering Research Board (SERB), Department of Science and Technology (DST), Government of India. A.K. acknowledges the support of the core research grant  CRG/2021/002455 and the  MATRICS grant MTR/2021/000350 from the SERB, DST, Government of India. 
A.D., M.K., and A.K. acknowledges support of the Department of Atomic Energy, Government of India, under Project No. 19P1112R\&D. This research was supported in part by the International Centre for Theoretical Sciences (ICTS) for participating in the program - Physics with Trapped Atoms, Molecules and Ions (code: ICTS/TAMIONs-2022/5) and Bangalore School on Statistical Physics - XII (code:  ICTS/bssp2021/6).
\end{acknowledgments}

\appendix
\section{Analytical forms for local density profile $n_i(t)$ and total occupation $N(t)$}
\label{app:scaling}

In this appendix, we present the details of the derivation of the analytical forms for local density profile $n_i(t)$ given in Eq.~\eqref{eq:ni_int_exact-m} and total occupation $N(t)$ given in Eq.~\eqref{eq:N_t_scaling-m}. We start with the equations of motion for the correlation function [Eq.~(\ref{eq:correlation})] which we recall below
\begin{eqnarray}
\label{eq:app_mod-corr-Lind}
    \frac{d C_{i,j}}{dt} &= i \, g\, (C_{i-1, j} \!-\! C_{i, j+1} \!+\! C_{i+1, j} \!-\! C_{i, j-1})\nonumber \\ 
    & -(\Gamma_L \mp \Gamma_G) (\delta_{i m} + \delta_{j m})\, C_{i, j} + 2\, \Gamma_G \delta_{m i} \delta_{m j} \nonumber \\
\end{eqnarray}
where $\mp$ stands for bosons and fermions, respectively. Note that $\Gamma_G$ is defined in Eq.~(\ref{eq:GammaG}) and $\Gamma_L$ is defined in Eq.~(\ref{eq:GammaL}). We will closely follow Ref.~\onlinecite{krapivsky2019free} and Ref.~\onlinecite{krapivsky2020free} to derive the analytical form for the density profile. For sake of brevity we define
\begin{equation}
\label{eq:gprime}
    \Gamma^{\prime} = \Gamma_L \mp \Gamma_G
\end{equation}
where $\mp$ stands for bosons and fermions, respectively. 
It is easy to see from Eq.~(\ref{eq:GammaG}) and Eq.~(\ref{eq:GammaL}) that the following inequality holds for both bosons and fermions, \begin{equation}
    \Gamma^{\prime} > 0.
\end{equation}
Therefore, Eq.~\eqref{eq:app_mod-corr-Lind} can be rewritten as 
\begin{eqnarray}
\label{eq:gp_app_mod-corr-Lind}
    \frac{d C_{i,j}}{dt} &= i \, g\, (C_{i-1, j} \!-\! C_{i, j+1} \!+\! C_{i+1, j} \!-\! C_{i, j-1})\nonumber \\ 
    & -\Gamma^{\prime} (\delta_{i m} + \delta_{j m})\, C_{i, j} + 2\, \Gamma_G \delta_{m i} \delta_{m j} 
\end{eqnarray}
with $\Gamma^{\prime} > 0$ always. Note that Eq.~\eqref{eq:gp_app_mod-corr-Lind} is an inhomogenous equation. Given that the lattice is initially in a vaccum, the following initial condition is satisfied 
\begin{equation}
\label{eq:ini-cond}
    C_{i,j}(t\!=\!0)=0.
\end{equation}
In order to solve Eq.~\eqref{eq:gp_app_mod-corr-Lind} along with the initial condition in Eq.~(\ref{eq:ini-cond}), we consider the following auxiliary problem. We will first solve Eq.~\eqref{eq:gp_app_mod-corr-Lind} without the in-homogenous piece $2\Gamma_G \delta_{m i} \delta_{m j}$. Let us write down the homogeneous equation as 
\begin{eqnarray}
\label{eq:h_gp_app_mod-corr-Lind}
    \frac{d \tilde{C}_{i,j}}{d\tau} &= i \, g\, (\tilde{C}_{i-1, j} \!-\! \tilde{C}_{i, j+1} \!+\! \tilde{C}_{i+1, j} \!-\! \tilde{C}_{i, j-1})\nonumber \\ 
    & -\Gamma^{\prime} (\delta_{i m} + \delta_{j m})\, \tilde{C}_{i, j}
\end{eqnarray}
where the symbol tilde indicates an auxiliary function satisfying the homogeneous equation and we have used the symbol $\tau$ to differentiate the time variable with that of the actual problem. Closely following Ref.~\onlinecite{krapivsky2019free} and Ref.~\onlinecite{krapivsky2020free}, we make the following ansatz, 
\begin{equation}
\label{eq:ansatz}
\tilde{C}_{i,j} (\tau) = \tilde{S}_i (\tau) \tilde{S}_j^*(\tau).
\end{equation}
Plugging in the ansatz given in Eq.~\eqref{eq:ansatz} into Eq.~\eqref{eq:h_gp_app_mod-corr-Lind}, we can show that $\tilde{S}_i (t)$ satisfies the following differential equation,
\begin{equation}
\label{eq:si_t}
    \frac{d \tilde{S}_i}{d\tau} = i g \big[ \tilde{S}_{i+1} + \tilde{S}_{i-1} \big] - \Gamma^{\prime} \, \delta_{im}\tilde{S}_i
\end{equation}
where the time dependence $\tau$ on $\tilde{S}_i(\tau)$ in Eq.~\eqref{eq:si_t} has been dropped for the sake of brevity. 

One can show that solving the original inhomogeneous differential equation in Eq.~(\ref{eq:gp_app_mod-corr-Lind}) along with the initial condition in Eq.~(\ref{eq:ini-cond}), can be achieved via solving the auxiliary homogeneous equation in Eq.~(\ref{eq:h_gp_app_mod-corr-Lind}) with the initial condition 
\begin{equation}
\label{eq:c_init}
    \tilde{C}_{i,j}(\tau=0) = \delta_{im}\delta_{jm}.
\end{equation}
This auxiliary initial condition [Eq.~\eqref{eq:c_init}] translates into
\begin{equation}
\label{eq:s_init}
    \tilde{S}_i(\tau=0) = \delta_{im}.
\end{equation}

Without loss of generality,  we henceforth consider the middle site to be at $m=0$. 
Furthermore, we take the lattice size $L$ to be infinity, i.e., the lattice is now extended from $-\infty$ to $+\infty$.
Since the system is expected to be symmetric about the $0$-th site, for simplicity, we consider only the positive side of the lattice chain in the analysis presented below. Now our goal is to analyse Eq.~\eqref{eq:si_t} along with initial condition given by Eq.~\eqref{eq:s_init}. One can solve Eq.~\eqref{eq:si_t} along with the initial condition Eq.~\eqref{eq:s_init} using a combination of Laplace and Fourier transformations \cite{krapivsky2019free} and the solution is given as 
\begin{eqnarray}
\label{eq:lap-F}
\tilde{S}_i(\tau) &=& J_i(2 \, g\, \tau) \nonumber \\
&-& \Gamma^{'} \, \int_{0}^{\tau} d\bar{t}\, e^{-\Gamma^{'} \bar{t}} \,\, \Big(\frac{\tau-\bar{t}}{\tau + \bar{t}}\Big)^{i/2} \, J_i \Big[ 2\, g\, \sqrt{\tau^2 - \bar{t}^2}\Big] \nonumber \\
\end{eqnarray}
where $J_i(z)$ denotes the Bessel function of first kind.  Note that, the local density at a particular site $i$ at time $t$ is given as,
\begin{equation}
\label{eq:ni_int_exact}
    n_i(t) = 2\, \Gamma_G \, \int_{0}^{ t} d\tau \, |\tilde{S}_{i}(\tau)|^2.
    \end{equation}
Notice that in order to simplify Eq.~\eqref{eq:ni_int_exact} one needs to use Eq.~\eqref{eq:lap-F} which itself has an integral, thereby making the simplification of $n_i(t)$ in Eq.~\eqref{eq:ni_int_exact} complicated. Interestingly, it turns out that $n_i(t)$ can admit an interesting scaling form. To do so, let us take the following limits,
\begin{equation}
\label{eq:scaling_limits}
    i \to \infty, \quad t \to \infty,  \quad \nu = \frac{i}{2 \, g \,t}  \sim O(1).
\end{equation}
where $\nu \sim O(1)$ is the scaled variable that will be used later. Owing to the scaling limit described in Eq.~\eqref{eq:scaling_limits}, the upper limit of the integral can be set to infinity. Moreover,  the contribution this integral in Eq.~\eqref{eq:ni_int_exact} largely comes when the integrand is evaluated at large $\tau$. This can be checked numerically although it is not entirely obvious from Eq.~\eqref{eq:lap-F}. Therefore it is justified to simplify Eq.~\eqref{eq:lap-F} in the large $\tau$ limit. In order to do so we use the following relation that holds for large $\tau$. 
\begin{equation}
\label{eq:scaletau}
\Big(\frac{\tau-\bar{t}}{\tau + \bar{t}}\Big)^{i/2} \approx e^{-\frac{i \, \bar{t}}{\tau}}.
\end{equation}
Using Eq.~\eqref{eq:scaletau} in Eq.~\eqref{eq:lap-F} we obtain
\begin{equation}
\label{eq:scaledS}
    \tilde{S}_i(\tau) = \frac{i\, J_i(2 \, g \, \tau)}{i + \, \tau \Gamma^{'}}.
\end{equation}
Using the simplified form of $\tilde{S}_i(\tau)$ in Eq.~\eqref{eq:scaledS}, the local density at a particular site is given as
\begin{eqnarray}
\label{eq:ni_int}
    n_i(t)&=& 2\, \Gamma_G \, \int_{0}^{t} d\tau \, |\tilde{S}_i(\tau)|^2 \nonumber \\
    &=& 2\, \Gamma_G \, \int_{0}^{t} d\tau \, \frac{i^2 \big[  J_i(2 \, g \, \tau)\big]^2}{(i+\tau\,\Gamma^{'})^2}
\end{eqnarray}
where we recall that $\Gamma_G$ is defined in Eq.~\eqref{eq:GammaG} and $\tilde{S}_i(\tau)$ is given in Eq.~\eqref{eq:scaledS}. Note that $i$ in Eq.~\eqref{eq:ni_int} stands for lattice index. Eq.~\eqref{eq:ni_int} is a compact analytical expression for the local density profile under the condition given in Eq.~\eqref{eq:scaling_limits}. 
We now proceed to analytically derive the scaling form. To do so, we need to use the appropriate asymptotic expansion for the Bessel function that appears in Eq.~\eqref{eq:ni_int}. Now making a change of variable $\tau =t\,s$ and recalling $i = 2 \,g \,\nu \, t$, we rewrite Eq.~\eqref{eq:ni_int} as
\begin{equation}
\label{re-scaled-n}
   n_i(t)= 8\,\Gamma_G \,g^2\, \nu^2 \, t \int_{0}^{1} ds \, \frac{\big[J_{2 g \nu t} (2 \, g \, t\, s)\big]^2}{(2\, g\, \nu +s\,\Gamma^{'})^2}.
\end{equation}
To facilitate the implementation of the asymptotic form of Bessel function, it is convenient to introduce 
\begin{equation}
\mu=2 \, g\, \nu\, t, \quad  {\rm and} \, \quad z=\frac{s}{\nu}
\end{equation}
which simplifies Eq.~\eqref{re-scaled-n} as,
\begin{equation}
\label{new-re-scaled-n}
   n_i(t)= 8\,\Gamma_G \,g^2\, \nu\, t \int_{0}^{\frac{1}{\nu}} dz \, \frac{\big[J_{\mu} (\mu \, z)\big]^2}{(2\, g + z\,\Gamma^{'})^2}.
\end{equation}
In the large $\mu$ limit (for a fixed $z$) the asymptotic expansion of $J_{\mu}(\mu z)$ is given by \cite{abramowitz1972abramowitz,NIST:DLMF}
\begin{equation}
    J_{\mu}(\mu \, z) \sim \left(\frac{4 \, \zeta(z)}{1-z^2}\right)^{\frac{1}{4}} \, \frac{1}{2 \sqrt{\mu \, \pi}} \, \frac{1}{|\zeta(z)|^\frac{1}{4}} \, \mathcal{T}_{\mu}(\mu \, z) 
    \label{asym-form-1}
\end{equation}
where $\mathcal{T}_{\mu}(\mu \, z)$ is given by
\begin{widetext}
\begin{equation}
\mathcal{T}_{\mu}(\mu \, z)=
\begin{cases}
\exp\left[-\frac{2}{3} \mu \, \zeta(z)^{\frac{3}{2}}\right] & \text{for} \quad  0 < z \leq 1 \\
2 \cos\left[\frac{2}{3} \mu \, \left(-\zeta(z)\right)^{\frac{3}{2}} - \frac{\pi}{4}\right] & \text{for} \quad 1 < z \leq \infty 
\end{cases}
\label{asym-form-2}
\end{equation}
with 
\begin{equation}
\label{zeta-z}
    \zeta(z)= 
\begin{cases}
\left(\frac{3}{2}\right)^{2/3} \left(\log
   \left(\frac{\sqrt{z^2+1}+1}{z}\right)-\sqrt{1-z^2}\right)^{2/3} & \text{for} \quad  0 < z \leq 1 \\
   -\left(\frac{3}{2}\right)^{2/3} \left(\sqrt{z^2-1}-\sec ^{-1}(z)\right)^{2/3} & \text{for} \quad 1 < z \leq \infty 
\end{cases}
\end{equation}
We split Eq.~\eqref{new-re-scaled-n} as follows
\begin{equation}
\label{new-re-scaled-n-n}
   n_i(t) = 8\,\Gamma_G \,g^2\, \nu\, t \Bigg[\int_{0}^1 dz \, \frac{\big[J_{\mu} (\mu \, z)\big]^2}{(2\, g + z\,\Gamma^{'})^2} 
   +\int_{1}^{\frac{1}{\nu}} dz \, \frac{\big[J_{\mu} (\mu \, z)\big]^2}{(2\, g + z\,\Gamma^{'})^2}\Bigg].
\end{equation}
\end{widetext}
To evaluate Eq.~\eqref{new-re-scaled-n-n} we use the appropriate forms in Eqs.\eqref{asym-form-1}-\eqref{zeta-z} depending on the range of integration over $z$. It is easy to notice that the first integral in Eq.~\eqref{new-re-scaled-n-n} is exponentially suppressed in $\mu$. Hence keeping only the second term in Eq.~\eqref{new-re-scaled-n-n} and performing some manipulations, we get 
\begin{equation}
     n_i(t) = \frac{2\,\Gamma_G \,g}{\pi}  \,\int_{1}^{\frac{1}{\nu}} dz\, \frac{1}{\sqrt{z^2-1}} \, \frac{\mathcal{T}^2_{\mu}(\mu \, z)}{(2\, g + z\,\Gamma^{'})^2}. 
\end{equation}
We now use the expression for $T_\mu(\mu z)$ from Eq.~\eqref{asym-form-2} and get 
\begin{equation}
     n_i(t) = \frac{8\,\Gamma_G \,g}{\pi}  \,\int_{1}^{\frac{1}{\nu}} dz\, \, \frac{\cos^2\left[\frac{2}{3} \mu \, \left(-\zeta(z)\right)^{\frac{3}{2}} - \frac{\pi}{4}\right]}{(2\, g + z\,\Gamma^{'})^2 \, \sqrt{z^2-1}}
\end{equation}
For large $\mu$, the cosine-squared term in the numerator is highly oscillatory and therefore can be approximated by $1/2$. We finally obtain the following scaling form for the local density profile $n_i(t)$, 
\begin{eqnarray}
     n_i(t) &=& \Phi\left(\frac{i}{2 g t}\right), \quad {\text{where}} \nonumber \\
         \Phi(\nu) &=&  \frac{4\,\Gamma_G \,g}{\pi}  \,\int_{1}^{\frac{1}{\nu}} dz\, \frac{1}{\sqrt{z^2-1}} \, \frac{1}{(2\, g + z\,\Gamma^{'})^2}
         \label{eq:phi-scale-app}
\end{eqnarray}
as also given in the main text in Eq.~\eqref{eq:ni_scale-main-1} and Eq.~\eqref{eq:ni_scale-main-2}. Upon performing the integral in Eq.~\eqref{eq:phi-scale-app} we obtain
\begin{widetext}
\begin{equation}
\Phi(\nu) = \frac{{\tilde{g}} \, ( 1 + \nu \, \tilde{g}) \Big[\log (1 +  \nu \, \tilde{g} )-\log \left(\tilde{g}+\nu -\sqrt{\left(\tilde{g}^2-1\right) \left(1-\nu ^2\right)} \right)\Big]-\sqrt{\left(\tilde{g}^2-1\right) \left(1-\nu ^2\right)}}{\left(\tilde{g}^2-1\right)^{3/2} (\nu\, \tilde{g} +1)}
, \quad 0 < \nu < 1\end{equation}
\end{widetext}
which is given in Eq.~\eqref{eq:phi-scale-int} of the main text. Here we introduced the dimensionless variable  $\tilde{g}$ as
\begin{equation}
\tilde{g}=\frac{2g}{\Gamma'} = \frac{4g}{J(0)}.
\end{equation}
Our result in Eq.~\eqref{eq:phi-scale-app} exactly coincides with the one obtained in Ref.~\cite{krapivsky2019free} where a different approach was used. 
In a similar fashion, the total occupation number $N(t)$ is given by
\begin{eqnarray}
N(t) &=& \sum_{i=-\infty}^{\infty} n_i(t) = 4\, g \,t \,\int_{0}^{1} d\nu \, \Phi(\nu) \nonumber \\
&=& \frac{16 \, \Gamma_G\, g^2 t}{\pi} \int_{0}^{1} d\nu \int_{1}^{\frac{1}{\nu}} dz\, \frac{1}{\sqrt{z^2-1}} \, \frac{1}{(2\, g + z\,\Gamma^{'})^2}. \nonumber \\
\label{eq:Nt_app}
\end{eqnarray}
We next perform an integral by parts in Eq.~\eqref{eq:Nt_app} which further simplifies $N(t)$ as,
\begin{eqnarray}
N(t) &=& \frac{4\,\Gamma_G \,\tilde{g}^2 t }{\pi}  \,\int_{1}^{\infty} \frac{dz}{z}\, \frac{1}{\sqrt{z^2-1}} \, \, \frac{1}{(\tilde{g} + z)^2} \nonumber \\
&=&-\frac{2\,\Gamma_G \,t }{\pi (1-\tilde{g}^2)} \,  \Big[2 \tilde{g}-\pi(1-\tilde{g}^2) \nonumber \\
&& \qquad \qquad +  2\big(1-2 \tilde{g}^2\big) \frac{\cos^{-1}\big(\tilde{g}\big)}{\sqrt{1-\tilde{g}^2}}\Big]
\end{eqnarray}
which matches with the expression in Eq.~\eqref{eq:N_t_scaling-m} of the main text. 

\bibliography{references}

\end{document}